\documentclass{article}

\usepackage{PRIMEarxiv}

\usepackage[utf8]{inputenc} 
\usepackage[T1]{fontenc}    
\usepackage{hyperref}       
\usepackage{url}            
\usepackage{booktabs}       
\usepackage{amsfonts}       
\usepackage{amssymb}
\usepackage{amsmath}
\usepackage{color}
\usepackage{xspace}
\usepackage{booktabs}
\usepackage{pifont}
\usepackage{nicefrac}       
\usepackage{microtype}      
\usepackage{lipsum}
\usepackage{fancyhdr}       
\usepackage{subcaption}

\usepackage{graphicx}       
\graphicspath{{media/}}    

\title{On the Performance and Memory Footprint of Distributed Training: An Empirical Study on Transformers}

\author{
  Zhengxian Lu$\dagger$ \\
  College of Computer Science\\
  Nankai University \\
  Tianjin, China\\
  \texttt{luzx@mail.nankai.edu.cn} \\
   \And
  Fangyu Wang$\dagger$ \\
  Zhejiang Lab \\
  Zhejiang, China \\
  \texttt{wangfy@zhejianglab.edu.cn} \\
  \And
  Zhiwei Xu \\
  Haihe Lab of ITAI\\
  Tianjin, China \\
  \And
  Fei Yang* \\
  Zhejiang Lab \\
  Zhejiang, China \\
  \texttt{yangf@zhejianglab.com} \\
  \And
  Tao Li* \\
 College of Computer Science\\
 Nankai University  \\
  Tianjin,China \\
  \texttt{litao@nankai.edu.cn} \\
}

\begin{document}
\maketitle

\begin{abstract}
Transformer models have emerged as potent solutions to a wide array of multidisciplinary challenges. The deployment of Transformer architectures is significantly hindered by their extensive computational and memory requirements, necessitating the reliance on advanced efficient distributed training methodologies. Prior research has delved into the performance bottlenecks associated with distributed training, aiming to unravel these bottlenecks and suggest optimization directions. However, such analyses often overlook three aspects unique to Transformer models: the specialized architecture, the dependency on various distributed strategies, and the requirement to balance computational and memory overhead.

This paper aims to bridge this gap by offering a comprehensive examination of the performance bottlenecks inherent in distributed training of Transformer models, leveraging both theoretical analysis and empirical investigation. We propose an analytical framework tailored to these unique aspects of Transformers, facilitating a holistic evaluation of model architectures, distributed strategies, and resource consumption. Based on this analytical framework, we conduct a comparative analysis of theoretical performances and further systematically explore how various distributed training strategies fare in real-world scenarios. Most of the experimental results can be well explained by the analytical outcomes derived from the analytical framework. Notably, our findings suggest an advantage of pipeline parallelism over data parallelism for Transformer models. Moreover, we shed light on some unexpected outcomes, such as the potential for increased total memory overhead due to suboptimal model partitioning within pipeline parallelism. Additionally, we underscore the significance of communication block size and waiting time to further enhance performance.
\end{abstract}

\keywords{distributed training \and Transformer \and performance evaluation}

\section{Introduction}
The advent and rapid evolution of Large Language Models (LLMs) have significantly changed the landscape of natural language processing \cite{radford2018gpt, radford2019gpt2, achiam2023gpt, devlin2018bert, touvron2023llama}, leading to groundbreaking performances and spurring extensions into diverse research field such as computer vision \cite{dehghani2023scalingVIT}, multi-modal \cite{wang2023image}, and agents \cite{hong2023metagpt}. Despite their outstanding capabilities, efficient training of LLMs poses substantial challenges. The dominant performance of LLMs mainly stems from the Encoder or Decoder architectures, making conventional distributed training paradigms not fully applicable to Transformer model training. Besides, the memory consumption of training LLMs far exceeds the capacities of individual devices \cite{yuan2024llm}. Efficient distributed training schemes are required for Transformers to complete the memory-efficient training in an acceptable time \cite{Switch2022Fedus}. To address the training challenges of Transformers, data, tensor, and pipeline parallelism strategies have emerged to partition and distribute deep learning models across clusters \cite{narayanan2021efficient}. Memory optimization techniques, including ZeRO \cite{rajbhandari2020zero} and re-computation \cite{huang2019gpipe}, have been employed to alleviate memory constraints during training.

To examine the characteristics and suggest further optimization directions, recent studies have explored the performance of distributed training by evaluating aspects such as training throughput, communication time, and model accuracy across various deep learning workloads, and hardware platforms \cite{shams2017evaluation, shi2018dag, mahon2020performance, ko2021depth}. However, scant attention has been devoted to the challenges posed by Transformers. Most existing studies focused solely on data parallelism, leaving the performance of tensor and pipeline parallelism largely unexplored. Additionally, there is a lack of comprehensive analysis on how distributed training schemes affect both training time and memory requirements and how optimization techniques can reduce memory overhead.

This paper aims to provide guiding suggestions for training Transformers in clusters by comparing the performance and memory overhead of Transformers with conventional models under various distributed training strategies and optimization techniques. The scope of this paper includes data, tensor, and pipeline parallelism strategies, as well as two optimization techniques: ZeRO and re-computation. To ground our experimental findings and enhance the generalizability of the conclusions, we perform a theoretical analysis of the time and memory overhead. However, theoretical modeling for analyzing the differences between Transformers and other models presents several challenges. Firstly, the analysis should not be limited to a specific model but can reflect the performance differences brought about by different model architectures. Secondly, the modeling should analyze the performance of distributed strategies without predetermining specific model partitioning schemes. Finally, the theoretical analysis should decouple the complex training process to consider various factors such as communication and scheduling overheads.

To address these challenges, we introduce an analytical framework for evaluating distributed training performance. Our framework identifies four observations that facilitate the characterization of performance-impacting factors from model architectures, deduce model partitioning schemes of distributed strategies, and dissect training time and memory overhead into manageable parts for modeling. Based on these observations, our framework facilitates distributed training performance analysis by employing tensor sizes and computational time of operators, alongside operator and tensor layouts derived from distributed strategies. Guided by performance evaluation methods, we provide a theoretical analysis of training time and memory overhead. The theoretical analysis offers insights into training time by examining communication, computation, overlap of communication and computation, and scheduling overheads, as well as evaluating memory overhead via the sizes of model weights and activations. 

Our experiment further investigates training performance and resource consumption. Most of the experimental observations align with our theoretical analysis, indicating that Transformers are more amenable to pipeline parallelism than the commonly used data parallelism approach for other models. We also offer optimization directions to reduce the impact of communication time on scalability and highlight unexpected findings, such as increased total memory overhead due to suboptimal model partitioning in pipeline parallelism. Lastly, we examine the theoretical and practical effect of employing ZeRO and re-computation techniques for memory optimization. Our work addresses gaps in existing evaluations of distributed training, enabling a deeper comprehension of the intricacies involved in training Transformer models.

The following summarizes our contributions:
\begin{itemize}
    \item We introduce an analytical framework that facilitates the characteristics of model architectures, distributed strategies, and the performance impacts of time and memory overhead. Based on the observations, our framework proposes a systematic method for performance analysis, allowing for a comprehensive examination of models, distributed strategies, and performance in a cohesive manner.
    \item Based on the analytical framework, we establish a theoretical analysis for training time, encapsulating communication, computation overhead, overlap of communication operations, and scheduling inefficiencies. Moreover, the analysis offers insights into memory consumption by considering tensor sizes, ultimately highlighting the performance variances among different distributed strategies.
    \item Our investigation includes extensive experiments that not only validate the theoretical model but also delve into the discrepancies observed between theoretical predictions and practical outcomes. We summarize 13 significant remarks by these experiments.
    \item We delineate our findings across four dimensions: outcomes that align with theoretical predictions, deviations from expected results, future directions for enhancing the efficiency of distributed training, and the comparative performance analysis between Transformers and conventional models. We validate the effect of memory optimization techniques through both theoretical and empirical lenses.
\end{itemize}

The rest of the paper is organized as follows: Section~\ref{sec:pre} introduces the distributed strategies and optimization techniques; Section~\ref{sec:framework} details our proposed analytical framework; Section~\ref{sec:analysis} presents a thorough theoretical analysis via our framework; Section~\ref{sec:evaluation} describes our comprehensive large-scale experimental study; in Section~\ref{sec:discussion}, we discuss the implications of our findings and examine memory optimization techniques; Section~\ref{sec:related} reviews relevant literature; and finally, Section~\ref{sec:conclusion} concludes the paper.

\section{Background}
\subsection{Distributed parallel strategy}

\begin{figure}[t]
    \includegraphics[width=\linewidth]{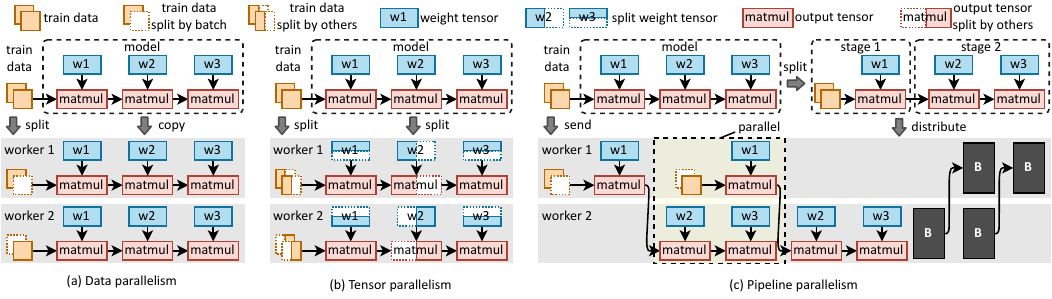}
    \centering
    \caption{An illustration of distributed parallel strategies, taking an example of training a 3-layer Multi-layer Perceptron (MLP). ``w1'', ``w2'', and ``w3'' are the weight tensor of {\ttfamily matmul} operators. The backpropagation in pipeline parallelism is shown as ``B'' to depict the scheduling.}
    \label{fig:strategies}
\end{figure}%

\textbf{Data parallelism} \cite{li2014scaling,li2020pytorch} involves distributing the workload across the batch dimension, where different computing devices perform identical computations on distinct subsets of the dataset, as depicted in Figure \ref{fig:strategies}a. Given the typically independent nature of sample computations, most operations in forward and backward passes do not necessitate communication, enabling data parallelism to achieve speed improvements across a broad spectrum of applications. Nonetheless, operations that require inter-sample statistical information, such as batch normalization \cite{ioffe2015batch}, necessitate either the accrual of statistics from local data subsets (potentially detracting from accuracy) or the introduction of additional communication overhead for global information exchange.

\textbf{Tensor parallelism} \cite{dryden2019improving,dean2012large}, also referred to as model parallelism \cite{Verbraeken2020survey} or horizontal parallelism \cite{Santhanam2021DistIR}, allocates computational workloads across devices by partitioning the weight or activation tensors of operators along one or more dimensions, as showcased in Figure \ref{fig:strategies}b. Implementing tensor parallelism requires devising partitioning schemes for each operator within a model, typically a task for domain experts. For instance, the Megatron-LM approach \cite{shoeybi2019megatron} proposes a specific tensor parallelism scheme for training Transformer models.

\textbf{Pipeline parallelism} \cite{huang2019gpipe,harlap2018pipedream,liu2022funcpipe} segments a model into several stages and distributes them across different devices. These stages are then executed in sequence, as illustrated in Figure \ref{fig:strategies}c. Communication overhead arises when there is a need to transfer activation tensors between operators spanning different stages. To enhance hardware utilization, pipeline parallelism introduces the concept of dividing a single batch into multiple micro-batches, thereby enabling parallel processing of stages through pipeline scheduling. Methods such as GPipe \cite{huang2019gpipe} utilize synchronous pipeline scheduling, incorporating a flush operation post the processing of each full batch to maintain consistency in training results. Conversely, asynchronous pipeline parallelism approaches, like the one proposed in \cite{harlap2018pipedream}, forgo the flush operation to achieve higher throughput in steady states, though at the risk of compromising on the assurance of training convergence. The scope of this paper is synchronous pipeline scheduling.

\subsection{Memory optimization technique}

\begin{figure}[t]
    \centering
    \begin{subfigure}[b]{0.49\linewidth}
        \includegraphics[width=\textwidth]{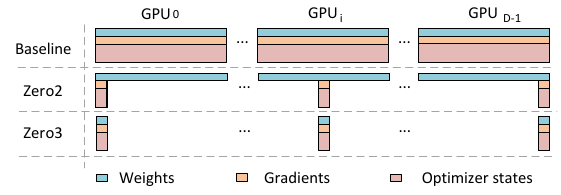}
        \caption{ZeRO \cite{rajbhandari2020zero}}
        \label{fig:zero}
    \end{subfigure}
    \hfill
    \begin{subfigure}[b]{0.49\linewidth}
        \includegraphics[width=\textwidth]{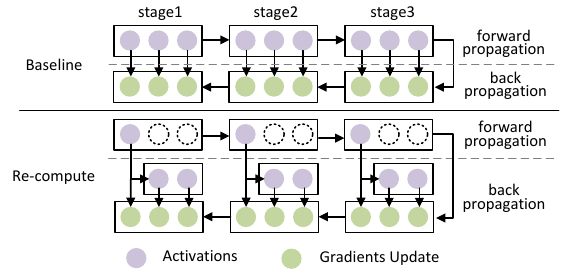}
        \caption{Re-computation}
        \label{fig:re-computation}
    \end{subfigure}
    \caption{An illustration of ZeRO and re-computation.}
\end{figure}

\textbf{ZeRO} \cite{rajbhandari2020zero} is designed to distribute optimizer states, gradients, and weights across different devices and perform collective communication when needed to reduce memory usage, as shown in Figure \ref{fig:zero}. In ZeRO-2, gradients and optimizer states are divided, allowing each device to store and compute a portion of the global gradients and optimizer states. ZeRO-3 extends this concept by also partitioning model weights, so each device only stores a portion of the weights and broadcasts them when required during forward and backward computations.

\textbf{Re-computation}, also known as activation checkpointing \cite{huang2019gpipe,narayanan2021efficient}, represents a strategy to trade computation resources for memory space. Re-computation involves retaining only a subset of the activation tensors and discarding the rest during forward propagation. Before backpropagation, the stored input tensors are employed to recompute the discarded activations, as demonstrated in Figure \ref{fig:re-computation}. When it is used in conjunction with the pipeline parallelism strategy, re-computation only stores the input tensors of the first operator for each stage while discarding all other activation tensors within the stages.

\section{Analytical framework}\label{sec:framework}

\subsection{Preliminary observations}

\textbf{Observation 1: Tensor properties cause the performance gap between different models.}
Current research suggests a hypothesis that the computation of the operators can be evenly divided to all participating devices \cite{zheng2022alpa,wang2019supporting}. Consequently, the gap of additional computational overhead for distributed training between different model architectures is modest. On the other hand, the relationship between the sizes of weight and activation tensors is decided by model architectures. For instance, for the input of a single sample, the size of the output activation (about 3.06MB) in the first convolutional layer of ResNet-50 \cite{he2016deep} is considerably larger than the weight (0.04MB). In contrast, the size of the weights and the output activations in the Encoder of Bert-base \cite{devlin2018bert} are comparable, being 27.04MB and 8.63MB respectively. The size of the tensor determines the communication volume, which in turn impacts the communication overhead.

\textbf{Observation 2: Operator and tensor layouts represent distributed strategies.}
A distributed strategy is a partitioning scheme of all the operators and tensors. Operators can be either assigned so that each device is exclusive to the computation of different operators (Figure \ref{fig:strategies}c), or be split across devices in a specific tensor partitioning scheme (Figure \ref{fig:strategies}a and \ref{fig:strategies}b). We use operator layout to represent the partitioning of operators and denote them as \textit{Exclusive(E)-layout} and \textit{Shared(S)-layout}, respectively. Furthermore, a tensor in an operator can be either replicated in all devices (weight tensors in Figure \ref{fig:strategies}a), or be partitioned into devices along one or more dimensions (weight tensors in Figure \ref{fig:strategies}b). We use tensor layout to represent the tensor partitioning scheme and denote them as \textit{Replicated(R)-layout} and \textit{Partitioned(P)-layout}, respectively. The operator and tensor layouts represent schemes for dividing computational workloads, leading to various patterns of communication and computation scheduling.

\textbf{Observation 3: The training time can be decoupled into four parts.}
Communication time, device computation time, device idle time caused by scheduling, and the overlapping time of computation and communication can be used to model training time. Denoting these parts as $T_{cm}$, $T_{cp}$, $T_{sd}$, and $T_{ol}$, respectively, the training time for one iteration on the device $d$ can be expressed as:
\begin{equation}
    T^d=T_{cm}^d+T_{cp}^d-T_{ol}^d+T_{sd}^d
    \enskip.
\end{equation}
By breaking down the training time into these components, we facilitate a more concise modeling of performance, enhancing the comprehensibility of the analysis.

\textbf{Observation 4: Memory overhead is related to tensor size.}
The memory demand comes from the storage of tensors and intermediate data during model creation, forward computing, backpropagation, and weight updating. Let $w_p$ and $a_p$ denote the sizes of weight and input tensors in operator $p$, respectively. The memory requirements can be summarized as follows: 1) model creation spends $\sum_{p\in\mathcal{P}} w_p$ to store weight tensors; 2) Forward computing requires $\sum_{p\in\mathcal{P}} a_p$ to store activations tensors; 3) backpropagation occupies $\sum_{p\in\mathcal{P}} w_p$ for gradient storage; and 4) weight updating utilizes $\alpha\sum_{p\in\mathcal{P}} w_p$ for optimizer states, where $\alpha$ depends on the type of optimizer. For instance, Adam optimizer \cite{kingma2014adam} incurs a memory requirement of $2\sum_p w_p$ to accommodate first-order and second-order moments, implying $\alpha=2$. Therefore, the total memory space required for model training $M$ can be represented using the sizes of weights and activations as:
\begin{equation} \label{eq:memory0}
    M = (\alpha+2)\sum_{p\in\mathcal{P}} w_p+\sum_{p\in\mathcal{P}} a_p+\gamma
    \enskip,
\end{equation}
where $\gamma$ represents the temporary memory footprint for the computation workspace.

\subsection{Methodology}

\begin{figure}[t]
    \centering
    \includegraphics[width=.6\linewidth]{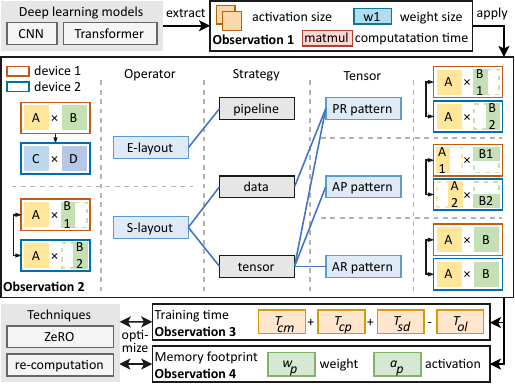}
    \caption{An illustration shows how our framework analyzes the performance and memory consumption of distributed training.}
    \label{fig:framework}
\end{figure}

Based on the four observations, our framework analyzes training performance as in Figure \ref{fig:framework}. Following \textbf{Observation 1}, we use the sizes of the activations and weights of the deep learning models, as well as the forward and backward computation times of the operators to represent the features of model architectures. Notably, our focus is on analyzing the performance differences between models. While computation time is crucial in estimating the overall training time, it exhibits a narrow effect on the scalability of distributed training based on Observation 1. The estimation of operator computation time can rely on the computational workload of the operators and the computing capacity of the devices, and use models such as the roofline model \cite{williams2009roofline}. However, adopting more detailed modeling approaches will not impact the conclusions of our theoretical analysis.

Our analytical framework distills the model partitioning schemes of distributed strategies based on \textbf{Observation 2}. As shown in Figure \ref{fig:framework}, pipeline parallelism leverages \emph{E-layout} and does not have a tensor layout. Data parallelism and tensor parallelism belong to the \emph{S-layout}. In data parallelism, activation tensors of all the operators are in \emph{P-layout} (divided along the batch dimension), while weight tensors are in \emph{R-layout}. The tensors in tensor parallelism can be either in \emph{P-layout} or in \emph{R-layout}. For brevity, we refer to an operator as in \emph{AR pattern} if all weights and input tensors within an operator are in \emph{R-layout}. Similarly, \emph{AP pattern} means that the weights and input tensors of the operator are all in \emph{P-layout}, and \emph{PR pattern} represents that the tensors of the operator include both \emph{P-layout} and \emph{R-layout}.

We meticulously model the time and memory resources required for distributed training according to \textbf{Observation 3}. Our framework divides performance into communication $T_{cm}$, computation $T_{cp}$, overlapping between computation and communication $T_{ol}$, and scheduling overhead $T_{sd}$. Based on \textbf{Observation 4}, the memory consumption can be divided into weights, activations, gradients, and optimizer states, and can further be represented by the sizes of weights $w_p$ and activation tensors $a_p$.

Based on our framework, we conduct the theoretical analysis in five parts, including communication, computation, communication overlap, scheduling overhead, and memory footprint. Operators and tensor layouts are utilized in each part to differentiate between various distributed strategies. Tensor sizes, weight sizes, and computation times are applied throughout the modeling process, thereby providing characteristics of the model architecture. After theoretical analysis and experimental study, we validated the effects of memory optimization techniques using our framework and analytical results.

\section{Theoretical analysis}\label{sec:analysis}

\subsection{Communication}

Pipeline parallelism follows the \emph{E-layout}, causing P2P communication across adjacent stages. In pipeline parallelism, each stage receives the output tensors from the previous stage and transmits its output tensors to the subsequent stage ($2\times$). The total communication volume including forward and backward propagation ($2\times$) is: 
\begin{equation}
    V_{pipe}=\sum_{d=1}^DV_{pipe}^d=4\sum_{d=2}^{D}a_{p(d,1)}
    \label{eq:pipevolume}
    \enskip,
\end{equation}
where $a_p$ denotes the input tensor size of operator $p$ and $p(d,1)$ denotes the first operator of stage $d$. $D$ represents the number of devices in pipeline parallelism. Denoting $b$ as the bandwidth of devices, the average communication overhead can be presented as:
\begin{equation}
    T_{cm,pipe}^d=\frac{1}{bD}V_{pipe}=\frac{4}{bD}\sum_{i=2}^{D}a_{p(i,1)}
    \enskip.
\end{equation}

Data and tensor parallelism follow the \emph{S-layout} and employ collective communications to collaborative completion operator computations. Collective communication may occur before, during, and after operator computation.
(1) Before operator computation, tensor layout conversion may occur to satisfy the operand tensor layout required by the operator. For example, if the input tensor is \textit{P-layout} while the operator requires the input tensor in \textit{R-layout}, an all-gather communication will be performed before the operator computation.
(2) During operator computation, communication may occur to obtain the output tensor. For instance, after applying the \textit{AP pattern} to matrix multiplication $C=AB$, $C = [A_1, A_2][B_1, B_2]^T=\sum A_i B_i$ computation requires all-reduce collective communication to sum all the $A_i B_i$.
(3) After the back-propagation of operators with the \textit{PR pattern}, an all-reduce communication is required to aggregate the results of \textit{R-layout} operands.

Data parallelism adopts \textit{PR pattern} for all the operators in models. The input tensors for all operators are split along the batch dimension. Therefore, the communication volume before operator computation is $0$. Due to the independent calculation of different samples, the communication volume during operator computation is $0$. Since all the weights are \textit{R-layout}, data parallelism should aggregate all the weights after back-propagation. The communication volume of data parallelism is the sum of the three parts as follows:
\begin{equation}
    V_{data} =\sum_{d=1}^D\sum_{p\in\mathcal{P}}V_p^d = 2(D-1)\sum_{p\in\mathcal{P}}w_{p}
    \label{eq:datavolume}
    \enskip,
\end{equation}
where $\mathcal{P}$ denotes the set of operators, and $V_p^d$ represents the communication volume required by operator $p$ on device $d$. In tensor parallelism, the communication volume $V_{tensor}^{d}$ is the cumulative total of communication volumes resulting from the partitioning of different tensor patterns. 

Collective communication overhead includes the actual communication time and the waiting time for all devices to be ready. Denoting $h_{p}^d$ as the waiting time needed for collective communication of operator $p$ on device $d$, the collective communication time could be estimated as:
\begin{equation}
    T_{cm,coll}^{d} = \sum_{p\in\mathcal{P}}(\frac{V_p^d}{b}+h_{p}^d)
    \enskip.
\end{equation}

Conclusively, the average communication volume of pipeline parallelism is related to the activation size of a few operators at the stage boundaries. The communication volume of data and tensor parallelism depends on the tensor sizes of all operators. Besides, the waiting time is an additional overhead for the E-layout parallel strategies that primarily involve collective communication.

\subsection{Computation}

We denote the forward and backward computation time of operator $p$ as $f_p$ and $b_p$, respectively. For pipeline parallelism with the \emph{E-layout}, operators are assigned to different devices. If operator $p$ is on device $d$, the computation time of operator $p$ on device $d$, denoted as $f_p^d$ and $b_p^d$, equals $f_p$ and $b_p$. Otherwise, we have $f_p^d = b_p^d = 0$. Therefore, the average device computational time is as follows:
\begin{equation}
    T_{cp,pipe}^d = \frac{1}{D}\sum_{i=1}^D \sum_{p\in\mathcal{P}} (f_p^i+b_p^i) = \frac{1}{D} \sum_{p\in\mathcal{P}} (f_p+b_p)
    \enskip.
\end{equation}

For operators with AP and PR patterns, the computational workload is distributed to all devices with a \emph{P-layout}. Ideally, the average computation time per device is $f_p^d = f_p/D$ and $b_p^d = b_p/D$. For operators with the \emph{AR pattern} (denoting as $\mathcal{P}_{AR}$, $\mathcal{P}_{AR}\subseteq\mathcal{P}$), replicated computations are performed on each device, and we have $f_p^d = f_p$ and $b_p^d = b_p$. We can derive the average device computation time for data and tensor parallelism as follows:
\begin{equation}
    T_{cp,data}^d = \frac{1}{D} \sum_{p\in\mathcal{P}} (f_p+b_p)
    \enskip,
\end{equation}
\begin{equation}
    T_{cp,tensor}^d = \sum_{p\in(\mathcal{P}-\mathcal{P}_{AR})} \frac{f_p+b_p}{D} + \sum_{p\in\mathcal{P}_{AR}} (f_p+b_p)
    \enskip.
\end{equation}
However, data and tensor parallelism employ collective communications, which subsequently reduces computational resources, leading to a decline in computational capability and augmenting $f_p^d$ and $b_p^d$.

Hence, data and pipeline parallelism do not introduce additional computational time costs. The same is true for tensor parallelism when the computation time of AR pattern operators can be considered negligible.

\subsection{Communication overlapping}

For pipeline parallelism, the communication of the $i$th micro-batch can overlap with the computation of the $(i-1)$th and $(i+1)$th micro-batches. Let $t_{ol}(i,j) =\min(a_{p(i,1)}/b, \sum_pf_p^{j}) +\min(a_{p(i,1)}/b, \sum_pb_p^{j})$ represent the overlap time between the communication time about the activation $a_{p(i,1)}$ and the computation time of stage $j$. The average overlap time can be presented as:
\begin{equation}
    T_{ol,pipe}^d = \frac{(B-1)}{BD} \sum_{i=2}^D(t_{ol}(i,i-1)+t_{ol}(i,i))
    \enskip,
\end{equation}
where $B$ denotes the number of micro-batches. Particularly, assuming that the communication time is always less than the computation time, then we have $T_{ol,pipe}^d = (B-1)/B\cdot T_{cm,pipe}^d$. By increasing $B$, we can strive to maximize the overlap of communication overhead.

Data parallelism performs the all-reduce in a wait-free manner to overlap communication with other back-propagation computations practically. Ideally, only the communication of the last several operators, denoting as $\mathcal{P}_{lst}$, cannot be overlapped, and we have
\begin{equation}
    T_{ol,data}^d=\sum_{p\in (\mathcal{P}-\mathcal{P}_{lst})}(\frac{2(D-1)}{Db}w_p+h_p^d)
    \enskip.
\end{equation}
It can be observed that the upper limit of overlapping time for data parallelism is approximately equal to the communication time if the communication overhead of $\mathcal{P}_{lst}$ is negligible for the total communication time $T_{cm}^d$.

For tensor parallelism, communication before and during operator computation cannot overlap with subsequent computations because they depend on the completion of communication. For communication after operator computation, communication cannot overlap with subsequent back-propagation calculations except weight aggregation. It is worth noting that communication may also be overlapped by other unrelated computations depending on the model structure and scheduling scheme.

Therefore, communication in tensor parallelism is challenging to overlap with computation. On the other hand, the overlapping time of data and pipeline parallelism can ideally approach the communication time. Their practical overlapping times are related to $B$ and $\mathcal{P}_{lst}$, respectively.

\subsection{Scheduling overhead}

The scheduling in pipeline parallelism naturally contains bubbles, leading to scheduling overhead. In the case of the GPipe scheduling scheme, bubbles originate from the cold start, which refers to the waiting period for devices to sequentially compute each stage of the first or the last micro-batch and transmit activations between stages. Ideally, if the computation times of different stages are equal, the average scheduling overhead can be represented as:
\begin{equation}
    T_{sd,pipe}^d = \frac{D-1}{BD} \sum_{p\in \mathcal{P}} (f_p + b_p) + \frac{2(D-2)}{bBD}\sum_{i=2}^D a_{p(i,1)} 
    \enskip.
\end{equation}
If we disregard the communication time, we can obtain the ratio of bubble time to total time as:
\begin{equation}
    T_{bbl} = \frac{T_{sd,pipe}^d}{T_{cp,pipe}^d+T_{sd,pipe}^d} = \frac{D-1}{D+B-1}
    \enskip,
\end{equation}
which is consistent with the conclusion of \cite{huang2019gpipe}. It can be observed that to achieve scalability, $B$ should at least be linearly related to $D$. Nevertheless, workload imbalance between stages can lead to a blocking effect, causing other stages to wait for a longer time before they can begin their next micro-batch computation. In the limit condition, if the workload division is extremely unbalanced, then $T_{sd,pipe}^d$ will increase by approximately $(B-1)\times$.

In tensor parallelism, although load imbalances can lead to an accumulation of $T_{sd}$, workloads are typically evenly distributed across each device. Furthermore, data parallelism ensures that the workload on each device is identical, resulting in minimal scheduling overhead. 

Consequently, data and tensor parallelism have nominal scheduling overhead, while pipeline parallelism will lead to significant device idleness.

\subsection{Memory footprint}

In pipeline parallelism, the memory for weight and activation tensors is only stored on a single device. We denote $w_p^{d}$ and $a_p^{d}$ as the memory footprint of weight and activation tensors on device $d$, respectively. If operator $p$ is on device $d$, we have $w_p^{d} = w_p , a_p^{d} = n_da_p/B$, where $n_d \in \{0, 1, 2, \dots, B\}$ is the number of micro-batches required to store on device $d$ simultaneously. Otherwise, we have $w_p^{d}=a_p^{d} = 0$. 
For GPipe, $n_d = B$ holds. Based on the memory consumption model in (\ref{eq:memory0}), when $\gamma$ is negligible, the sum of memory footprints on all devices should remain unchanged with the increase of devices, satisfying: 
\begin{equation}
\begin{split}
     \sum_{d=1}^D M_{GPipe}^{d} & = \sum_{d=1}^D ((\alpha+2)\sum_{p\in\mathcal{P}} w_p^{d} +  \sum_{p\in\mathcal{P}} a_p^{d} ) \\
     & = (\alpha+2)\sum_{p\in\mathcal{P}} w_p +  \sum_{p\in\mathcal{P}} a_p 
     \label{eq:pipememory}
    \enskip.
\end{split}
\end{equation}
The 1F1B scheduling scheme \cite{harlap2018pipedream} further holds $n_d = \min\{1+S-d, B\}$ since the activation tensors are consumed by back-propagation earlier. The total memory footprint falls between $\sum_d M_{GPipe}^{d}$ and $(\alpha+2)\sum_p w_p$.

The memory footprint of data and tensor parallelism is related to the tensor layout. We define $\mathcal{P}_{wp}\subseteq\mathcal{P}$ as the set of operators with a weight in \textit{P-layout} and $\mathcal{P}_{ap}\subseteq\mathcal{P}$ as the set of operators with an activation tensor in \textit{P-layout}. We further denote $\mathcal{P}_{\overline{wp}} = \mathcal{P} - \mathcal{P}_{wp}$ and $\mathcal{P}_{\overline{ap}} = \mathcal{P} - \mathcal{P}_{ap}$. Operators in the \textit{AP pattern} are part of both $\mathcal{P}_{wp}$ and $\mathcal{P}_{ap}$. Conversely, operators in the \textit{AR pattern} are belonging to $\mathcal{P}_{\overline{wp}}$ and $\mathcal{P}_{\overline{ap}}$ instead. For the \textit{PR pattern}, operators may fall into $\mathcal{P}_{wp}$ or its complement $\mathcal{P}_{\overline{wp}}$, as well as into $\mathcal{P}_{\overline{ap}}$ or its complement $\mathcal{P}_{ap}$. The tensors under the \emph{P-layout} are distributed across multiple devices, so we have $\sum_d w_p^d = w_p, p\in \mathcal{P}_{wp}$ and $\sum_d a_p^d = a_p, p\in \mathcal{P}_{ap}$. The tensors under the \emph{R-layout} have a copy on each device, so we have $w_p^d = w_p, p\in \mathcal{P}_{\overline{wp}}$ and $a_p^d = a_p, p\in \mathcal{P}_{\overline{ap}}$. The total memory footprint is:
\begin{equation}
\begin{split}
    \sum_{d=1}^D M^d = & (\alpha + 2)[\sum_{p\in\mathcal{P}_{wp}} w_p + D \sum_{p\in \mathcal{P}_{\overline{wp}}} w_p] \\
    & + \sum_{p\in\mathcal{P}_{ap}} a_p + D\sum_{p\in \mathcal{P}_{\overline{ap}}} a_p 
    \enskip.
\end{split}
\end{equation}
Specifically, for data parallelism, all operators belong to $\mathcal{P}_{\overline{ap}}$ and $\mathcal{P}_{wp}$. The total memory allocation can be formulated as follows:
\begin{equation}
    \sum_{d=1}^D M_{data}^d = (\alpha + 2) D \sum_{p\in \mathcal{P}} w_p + \sum_{p\in\mathcal{P}} a_p 
    \label{eq:datamemory}\enskip.
\end{equation}

Accordingly, pipeline parallelism does not increase the memory footprint. Conversely, while data and tensor parallelism decrease memory usage on individual devices, they result in an increased total memory footprint across all devices. Furthermore, data parallelism achieves greater memory utilization efficiency when activation tensors account for a substantial portion of the memory footprint, which is prevalent in convolutional neural networks.

\section{Experimental evaluation}\label{sec:evaluation}

\subsection{Experiment settings}

\textbf{Workloads.} 
Since CNN is widely applied in the field of computer vision \cite{krizhevsky2017imagenet,ren2015faster,he2016deep}, we adopt CNN as the representative of traditional models. Specifically, we choose ResNet \cite{he2016deep} and Bert \cite{devlin2018bert} to represent the CNN and Transformer structure, respectively. We employ the ImageNet dataset \cite{deng2009imagenet} to assess the performance of the ResNet models, including ResNet50, ResNet101, and ResNet152. We use the Bert-base version of the Bert model, which comprises 12 Encoders. We train Bert-base for the Quora Question Pairs (QQP) task in the Glue dataset \cite{wang2018glue}, a binary classification task. Unless otherwise specified, the global batch size for the Bert model is $32\times$GPUs, and the ResNet models have a global batch size of $128\times$GPUs. 

\textbf{Partitioning schemes.}
The method of partitioning for pipeline parallelism involves dividing Bert according to the ``Encoder'', and ResNet according to the ``Bottleneck block''. Then, the execution time of each module is tested five times, and we determine a stage division scheme that minimizes the execution time of the longest-running stage.
In the tensor parallelism scheme, Bert employs the partitioning scheme from Megatron-LM \cite{shoeybi2019megatron}. Megatron-LM partitions the ``head'' dimension of the ``query'', ``key'', and ``value'' matrix multiplication and the corresponding weight matrices in the multi-head attention module, as well as the weight matrices of the ``feed-forward network'' module. Megatron-LM requires the aggregation of output tensors through all-reduce collective communication. ResNet utilizes the approach introduced in \cite{dryden2019improving}, which involves dividing activations, referred to as feature maps, by the ``height'' dimension and performing computations on different devices. When the size of the convolution kernel exceeds 1, an inter-device communication is generated to exchange border information of adjacent feature maps.

\textbf{Platforms.} 
Our experiments are based on PyTorch \cite{paszke2019pytorch} 1.10, evaluated on a cluster of 4 servers. Each server is equipped with 8 NVIDIA A100 GPUs, with each GPU having 80GB of memory. The eight GPUs in one server interconnect using NVLink. All servers are connected using four 200Gbps InfiniBand (IB) network cards.

\textbf{Metrics.}
We use the latest-generation performance analysis tool provided by NVIDIA, Nsight System, to gather experimental results. Nsight System is capable of monitoring the model training process and provides detailed information on various aspects such as CUDA API calling and execution, GPU status, and communication bandwidth. After running several training steps to bring the system into a stable state, we collect the experimental results based on the average of 100 training steps.

\subsection{Communication}

\begin{figure}[t]
    \centering
    \begin{subfigure}[b]{.49\linewidth}
        \includegraphics[width=\textwidth]{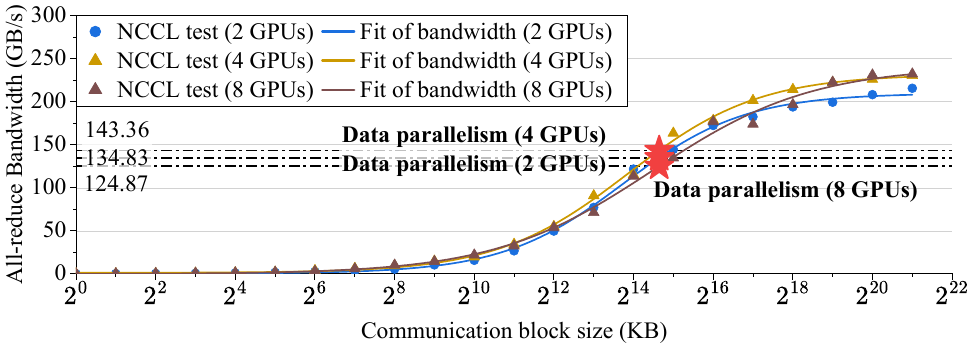}
        \caption{All reduce bandwidth}
    \end{subfigure}
    \hfill
    \begin{subfigure}[b]{.49\linewidth}
        \includegraphics[width=\textwidth]{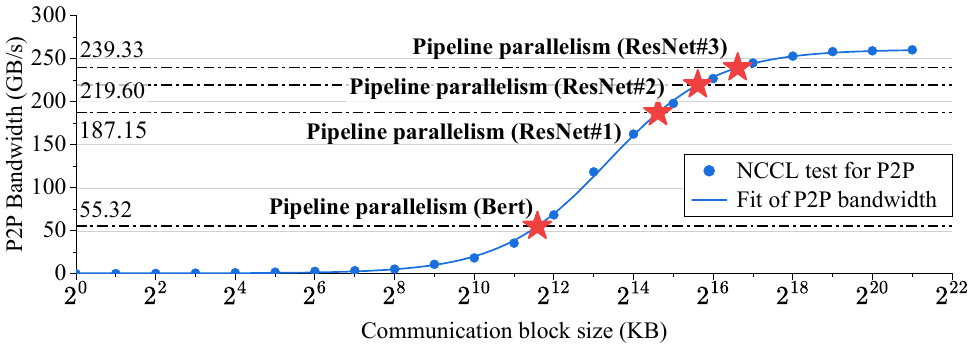}
        \caption{P2P bandwidth}
    \end{subfigure}
    \caption{The bandwidth measured by NCCL test (scatter plots) and fitted with sigmoidal curves (lines). ``ResNet\#1-\#3'' refers to the three sizes of feature maps in ResNet.}
    \label{fig:bandwidth-nccltest}
\end{figure}

We first compared the differences between predicted and measured communication time. According to (\ref{eq:pipevolume}) and (\ref{eq:datavolume}), we predicted the communication time for pipeline and data parallelism as representatives of P2P communication and collective communication, respectively. We measured the bandwidth of P2P communication and all-reduce communication under different communication block sizes using the NCCL test tool\footnote{\url{https://github.com/NVIDIA/nccl-tests}}, and fitted them with sigmoidal curves, as shown in Figure \ref{fig:bandwidth-nccltest}. The communication block size for data parallelism is set to 25MB in PyTorch by default, while the size for pipeline parallelism is related to the size of the activations of a single micro-batch. We marked these communication block sizes in Figure \ref{fig:bandwidth-nccltest}. It can be observed that due to the differences in communication methods and communication block sizes, data and pipeline parallelism have a variety of bandwidths. However, most of these bandwidths are far below the theoretical bandwidth of NVLink.

\textbf{Remark1} The sizes of most communication blocks fall within the range where the bandwidth significantly varies with the change in communication block size.

\begin{figure}[t]
    \centering
    \begin{subfigure}[b]{.49\linewidth}
        \includegraphics[width=\textwidth]{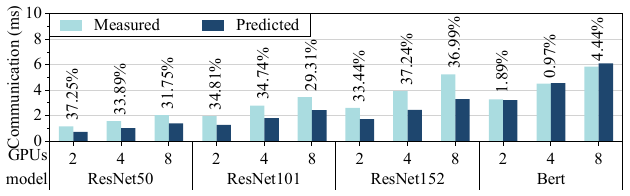}
        \caption{Data parallelism}
    \end{subfigure}
    \hfill
    \begin{subfigure}[b]{.49\linewidth}
        \includegraphics[width=\textwidth]{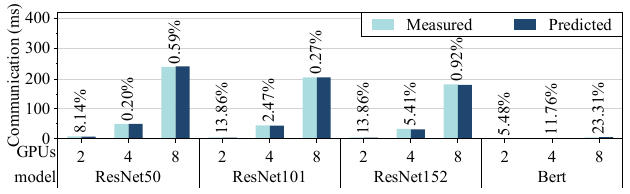}
        \caption{Pipeline parallelism}
    \end{subfigure}
    \caption{The prediction of communication time based on activation sizes and the fitted bandwidth. The percentage error is shown above each prediction.}
    \label{fig:commvolumepred}
\end{figure}

\begin{figure}[t]
    \centering
    \includegraphics[width=\linewidth]{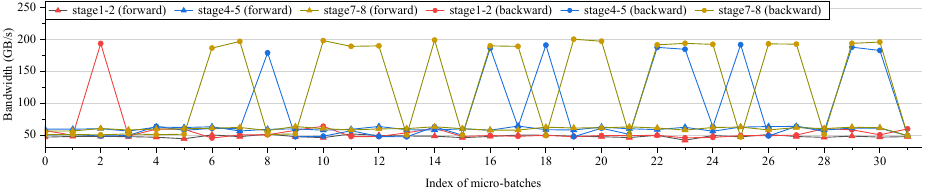}
    \caption{The communication bandwidth of 32 micro-batches in a single training step when using 8 GPUs to train the Bert in pipeline parallelism. Most communication bandwidth fluctuates around 55.32GB/s, with a small portion approaching 200GB/s.}
    \label{fig:bandwidth-exception}
\end{figure}

Based on these bandwidths, we predict the communication time for data and pipeline parallelism. The comparison between the predicted communication time and the measured time is shown in Figure \ref{fig:commvolumepred}. It can be observed that the measured communication time for Bert under data parallelism and ResNet under pipeline parallelism is close to our predictions. For data parallelism, the reason for the significant increase in measured time for ResNet compared to predictions is that the Batch Normalization \cite{ioffe2015batch} included in the ResNet architecture requires additional broadcast communication. For pipeline parallelism, the large prediction error for Bert is due to the bandwidth of a few P2P communications significantly exceeding our bandwidth predictions in Figure \ref{fig:bandwidth-nccltest}. Figure \ref{fig:bandwidth-exception} shows the forward and backward communication bandwidths for all micro-batches under 8 GPUs. Most bandwidths are close to our predictions (55.32GB/s), while a few bandwidths in the backward propagation approach are 200GB/s. The sudden increase in bandwidth might be caused by the short time intervals of consecutive P2P communications.

\textbf{Remark2} According to prior bandwidth information, it is possible to estimate the collective communication time of Transformers and P2P communication time based on the distributed strategy and tensor sizes.

\begin{figure}[t]
    \centering
    \begin{subfigure}[b]{0.49\linewidth}
        \includegraphics[width=\textwidth]{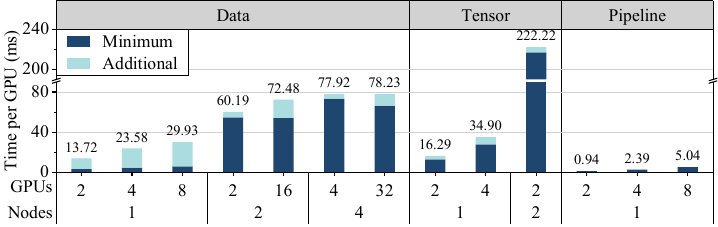}
        \caption{Bert}
    \end{subfigure}
    \hfill
    \begin{subfigure}[b]{0.49\linewidth}
        \includegraphics[width=\textwidth]{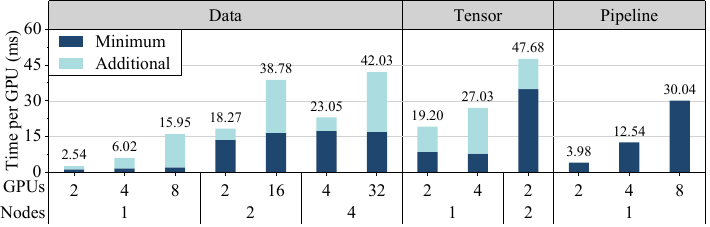}
        \caption{ResNet50}
    \end{subfigure}
    \hfill
    \begin{subfigure}[b]{0.49\linewidth}
        \includegraphics[width=\textwidth]{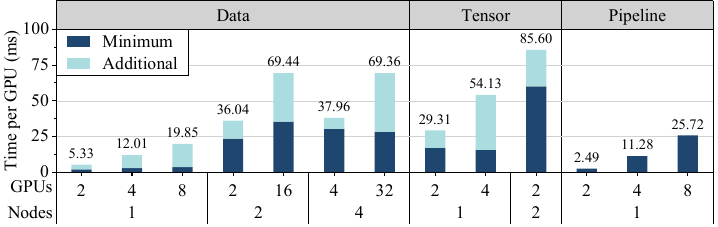}
        \caption{ResNet101}
    \end{subfigure}
    \hfill
    \begin{subfigure}[b]{0.49\linewidth}
        \includegraphics[width=\textwidth]{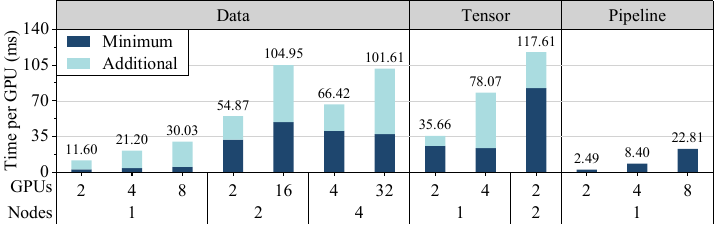}
        \caption{ResNet152}
    \end{subfigure}
    \caption{The average time per GPU for communication under various distributed strategy, node, and GPU configurations. ``Minimum'' represents the shortest communication time among GPUs during collective communication, and ``Additional'' indicates the communication time beyond the minimum value. The sum of communication time and waiting time is shown above each bar.}
    \label{fig:communication times}
\end{figure}

\begin{figure}[t]
    \centering
    \includegraphics[width=\linewidth]{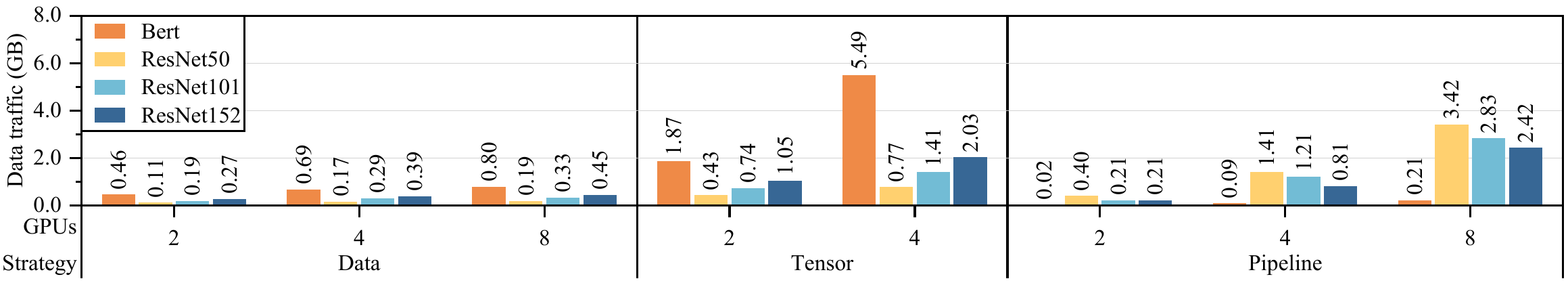}
    \caption{NVLink data traffic measured through Nsight System.}
    \label{fig:nvlink-commvolume}
\end{figure}

In Figure \ref{fig:communication times}, we present the communication times in scenarios with different numbers of nodes and GPUs, where ``Minimum'' represents the shortest communication time among GPUs during collective communication, and ``Additional'' indicates the communication time beyond the minimum value. ``Minimum'' can be approximately regarded as the actual communication time, while ``Additional'' can be seen as the time overhead for waiting between GPUs during collective communication.
We have three observations based on the experimental results. 1) In Bert, the communication time for pipeline parallelism is less than that for data parallelism, primarily because pipeline parallelism requires less communication volume. Conversely, in ResNet, the communication time for pipeline parallelism is longer than for data parallelism, especially as the number of devices increases. This is mainly due to the different data traffic during training, as shown in Figure \ref{fig:nvlink-commvolume}. 2) As the number of devices scales out across nodes, the main performance bottleneck in data parallelism for Transformers shifts from waiting time to actual communication time. Meanwhile, due to the impact of data preprocessing on CPUs, waiting times significantly affect the overall communication time in CNNs. This is primarily because the increase in the number of GPUs within a single node leads to intensified competition for CPU resources and hinders the in-time launch of kernel functions onto devices. 3) Scaling out the number of devices across nodes significantly increases the ``min'' part of communication time, especially when using tensor parallelism.

\textbf{Remark3} Employing data parallelism in CNNs is more suitable for use in bandwidth-limited environments. Conversely, in Transformers, pipeline parallelism has a significantly lower communication time. 

\textbf{Remark4} In collective communication, waiting occupies a large proportion of the communication time, especially in ResNet, or when scaling up within a node.

\textbf{Remark5} Regardless of the network, tensor parallelism requires high bandwidth between devices.

\subsection{Computation}

\begin{figure}[t]
    \centering
    \includegraphics[width=\linewidth]{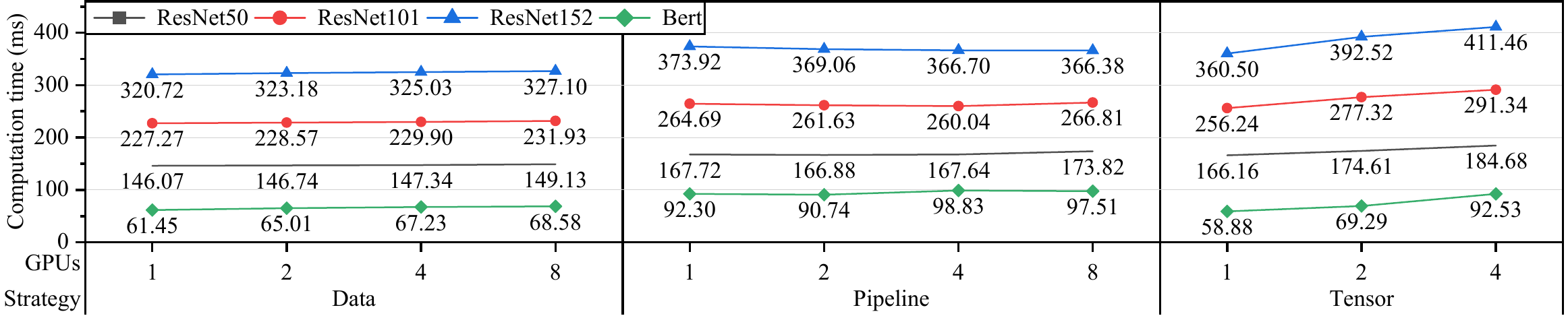}
    \caption{The relationship between the average computing time per GPU and the number of GPUs under three distributed strategies.}
    \label{fig:compScaleUp}
\end{figure}

Figure \ref{fig:compScaleUp} presents the device computation times for different distributed strategies. We can observe that there is no significant relation between the average computation time of pipeline parallelism and the number of devices involved in the computation. However, the computation time for data parallelism increases slightly with the number of devices, while tensor parallelism increases significantly. This difference arises from the fact that tensor parallelism runs several operators in \textit{AR pattern}, which leads to redundant computations, further increasing the computational workload and elevating the average computation time. Additionally, both data and tensor parallelism require collective communication. NCCL-based collective communication consumes computation resources such as cuda cores and stream multiprocessors, thereby resulting in decreased computational efficiency and, consequently, an increase in device computation time. We will further analyze this issue with communication overlapping. The change in computation time in pipeline parallelism may be due to the use of different kernel functions to calculate operators.

\textbf{Remark6} The computational efficiency of data and tensor parallelism decreases with the increase in the number of devices.

\subsection{Communication overlapping}

\begin{figure}[t]
    \centering
    \begin{subfigure}[b]{0.49\linewidth}
        \includegraphics[width=\textwidth]{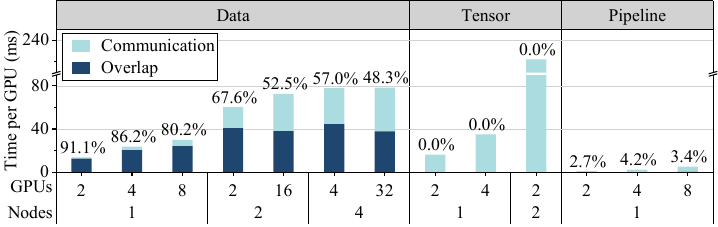}
        \caption{Bert}
    \end{subfigure}
    \hfill
    \begin{subfigure}[b]{0.49\linewidth}
        \includegraphics[width=\textwidth]{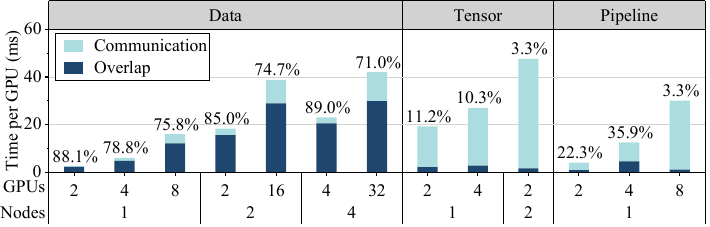}
        \caption{ResNet50}
    \end{subfigure}
    \hfill
    \begin{subfigure}[b]{0.49\linewidth}
        \includegraphics[width=\textwidth]{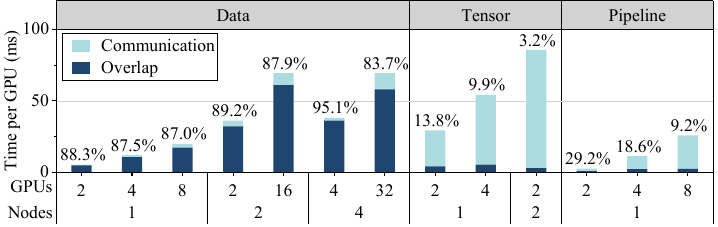}
        \caption{ResNet101}
    \end{subfigure}
    \hfill
    \begin{subfigure}[b]{0.49\linewidth}
        \includegraphics[width=\textwidth]{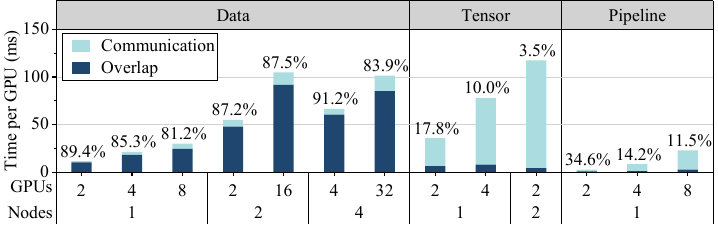}
        \caption{ResNet152}
    \end{subfigure}
    \caption{The communication time and overlap time among different models and distributed strategies. The percentage of overlap time to total communication time is shown above the bar.}
    \label{fig:overlapping}
\end{figure}

\begin{figure}[t]
    \centering
    \includegraphics[width=.6\linewidth]{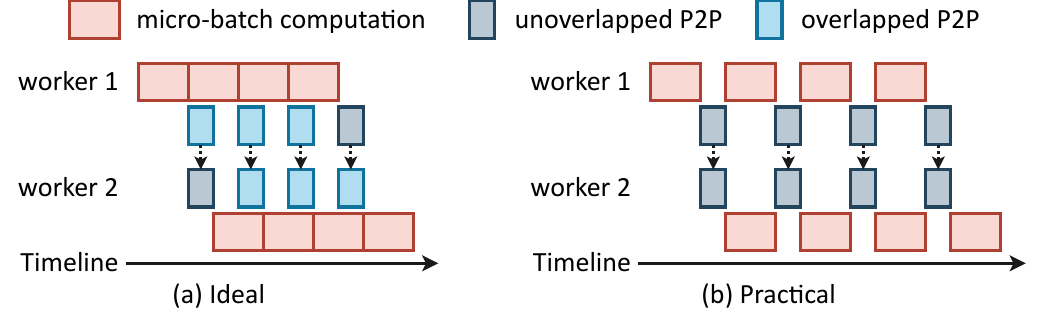}
    \caption{Comparison of ideal communication overlap in pipeline strategy (a) and practical communication overlap in PyTorch (b).}
    \label{fig:pytorchpipe}
\end{figure}

Figure \ref{fig:overlapping} illustrates the communication time and overlapping time. It can be observed that data and tensor parallelism exhibit higher communication ratios. Data parallelism also has a higher overlap ratio, while tensor parallelism has almost no communication overlap, especially for Bert. For pipeline parallelism, only a slight portion of communication time is overlapped. The reason for contradicting the analytical conclusion is that PyTorch blocks processes during forward propagation until all devices complete single micro-batch computation or P2P communication. As in Figure \ref{fig:pytorchpipe}, the frequent synchronizations prevent effective communication overlapping.

\textbf{Remark7} The overlap time of data and tensor parallelism is consistent with analytical results. However, an improper pipeline parallelism implementation can result in communication times not overlapping.

\begin{figure}[tb]
    \centering
    \includegraphics[width=\linewidth]{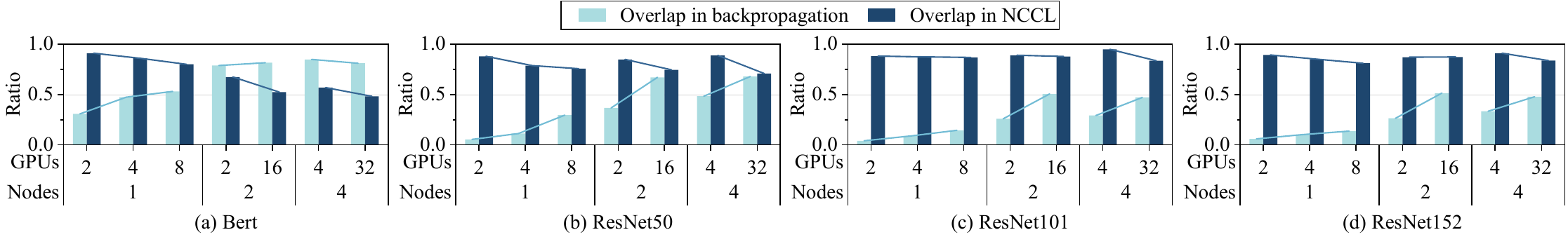}
    \caption{The proportion of overlap in backpropagation computation time and communication time.}
    \label{fig:overlapping_ratio}
\end{figure}

From Figure \ref{fig:overlapping}, we can also observe that the communication overlap ratio of Bert is lower than that of ResNet with multiple nodes under data parallelism. Figure \ref{fig:overlapping_ratio} further illustrates the proportion of computation and communication overlap in data parallelism. The term ``overlap in NCCL'' denotes the proportion of overlap time to the total communication time, whereas ``overlap in backpropagation'' signifies the proportion of overlap time to the computation time in backpropagation. These proportions represent whether the overlapping time reaches the upper limit of communication and computation time. With an increase in the number of nodes and GPUs, ResNet50 exhibits a relatively stable ``overlap in NCCL'', while ``overlap in backpropagation'' invariably remains below that of ``overlap in NCCL''. This implies the potential for overlapping more communication time. However, in Bert, as the number of devices expands, ``overlap in NCCL'' decreases, while ``overlap in backpropagation'' surpasses ``overlap in NCCL'' and reaches a saturation point. This indicates that the communication that can be concealed by computation reaches its limit, and communication will become the bottleneck for expanding the training scale.

\textbf{Remark8} Transformer exhibits overlap characteristics that differ from those of CNNs in data parallelism. There is a significant amount of communication overhead that cannot be overlapped.

\begin{figure}[t]
    \centering
    \begin{subfigure}[b]{0.49\linewidth}
        \includegraphics[width=\textwidth]{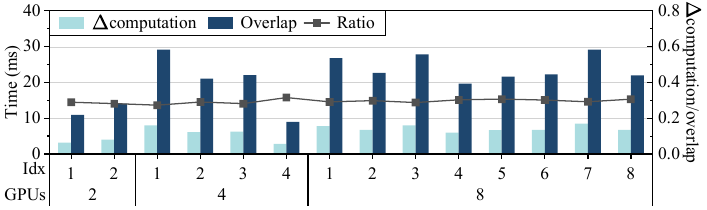}
        \caption{Bert}
    \end{subfigure}
    \hfill
    \begin{subfigure}[b]{0.49\linewidth}
        \includegraphics[width=\textwidth]{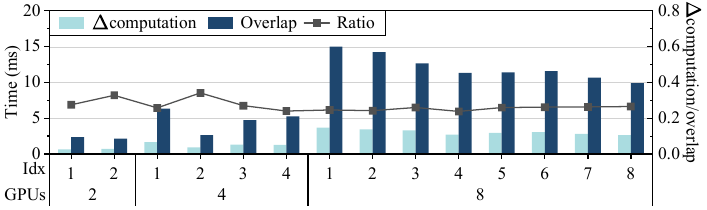}
        \caption{ResNet50}
    \end{subfigure}
    \hfill
    \begin{subfigure}[b]{0.49\linewidth}
        \includegraphics[width=\textwidth]{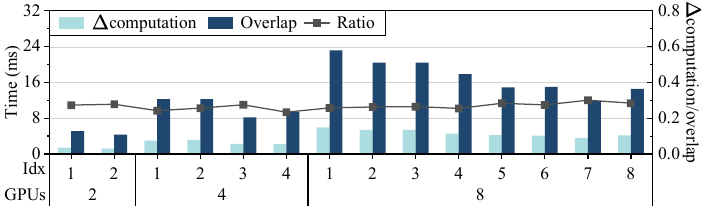}
        \caption{ResNet101}
    \end{subfigure}
    \hfill
    \begin{subfigure}[b]{0.49\linewidth}
        \includegraphics[width=\textwidth]{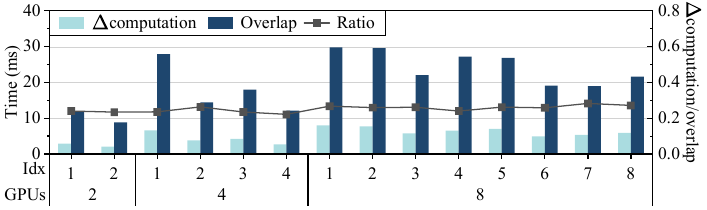}
        \caption{ResNet152}
    \end{subfigure}
    \caption{Under data parallelism with different numbers of GPUs, the relationship between the increase in computation time of each GPU compared to single-card training and the overlap time. In the X-axis, "GPUs" and "Idx" respectively represent the number of GPUs and the index of GPUs in distributed training. Line charts represent the ratio of the increase in computation time to overlap time.}
    \label{fig:ncclimpact}
\end{figure}

We also observed that overlapping communication with NCCL leads to an increase in the execution time of kernel functions. Figure \ref{fig:ncclimpact} shows the relationship between the increase in computation time ($\Delta$computation in the figure) for each GPU and the overlapping time when using 2, 4, and 8 GPUs for data-parallel distributed training, compared to the computation time of training with a single GPU. The line chart in the figure depicts the ratio of the increase in computation time to the overlapping time. It can be observed that the ratio of the increase in computation time to the overlapping time remains stable at around 0.3. This is mainly because the execution of NCCL requires CUDA cores. The decrease in computing resources causes the computation time to increase proportionally.

\textbf{Remark9} The increase in computation time for data parallelism stems from NCCL occupying computational resources.

\subsection{Scheduling overhead}

\begin{figure}[t]
    \centering
    \includegraphics[width=\linewidth]{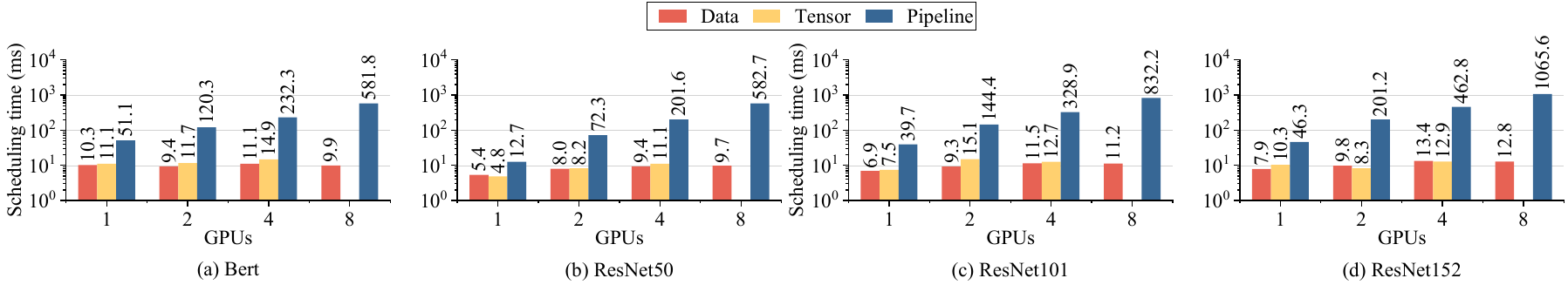}
    \caption{Comparison of scheduling time in models and distributed strategies.}
    \label{fig:scheduling}
\end{figure}

Figure \ref{fig:scheduling} shows the idle time across different GPUs for various distributed strategies. It can be observed that data parallelism and tensor parallelism exhibit little idle time. However, pipeline parallelism is characterized by a significantly increased $T_{sche}$, attributed to its inherent scheduling overhead and the frequent synchronization of pipeline parallelism implementation in PyTorch. As shown in Figure \ref{fig:pytorchpipe}, the micro-batch computation is launched only after the preceding micro-batch computation is completed on all devices, leading to prolonged device idle time.

\textbf{Remark10} Consistent with the theoretical analysis, pipeline parallelism will bring significant scheduling overhead, thereby heavily reducing the training performance.

\subsection{Memory footprint}

\begin{figure}[t]
    \centering
    \includegraphics[width=\textwidth]{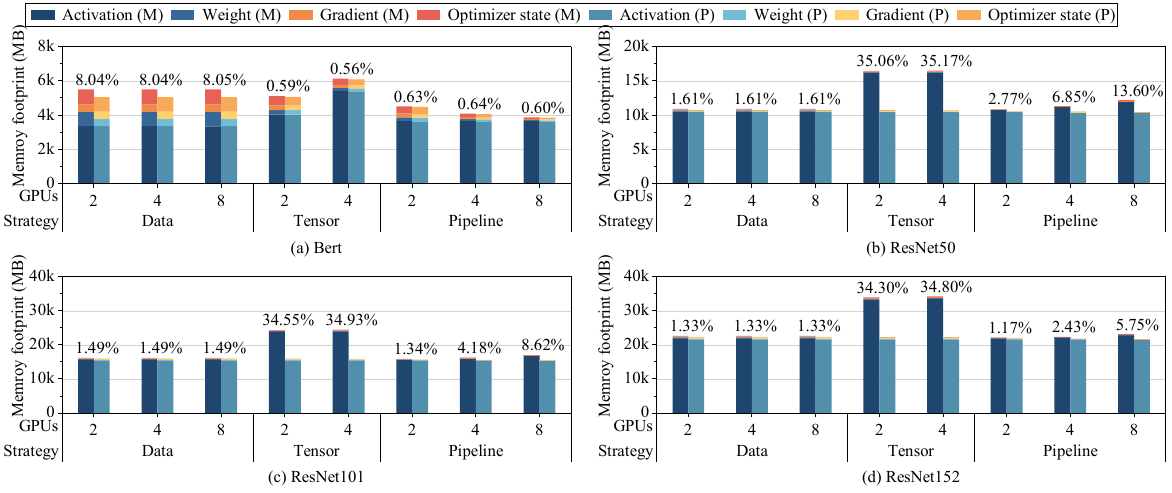}
    \caption{The comparison of measured (`M') and predicted (`P') memory footprint. Memory is further divided into weights, activations, gradients, and optimizer states. The percentage error is shown above each prediction.}
    \label{fig:predmem}
\end{figure}

Figure \ref{fig:predmem} shows a comparison of the average memory usage per GPU among different strategies and models. Memory footprint is further divided into weights, activations, gradients, and optimizer states. The predicted memory footprint is also depicted. We can observe that some of the measured values are close to the predicted values, especially for tensor and pipeline parallelism in Bert. The memory overhead for data parallelism remains constant, which is also consistent with our theoretical analysis results. Compared to ResNet, the memory consumption of weights, gradients, and optimizer states in Bert is non-negligible. Therefore, pipeline parallelism can reduce the total memory footprint more efficiently.

\textbf{Remark11} Pipeline parallelism is most effective in reducing the memory overhead of Transformers, as it simultaneously partitions both the activations and the weights.

\begin{figure}[t]
    \centering
    \includegraphics[width=.5\linewidth]{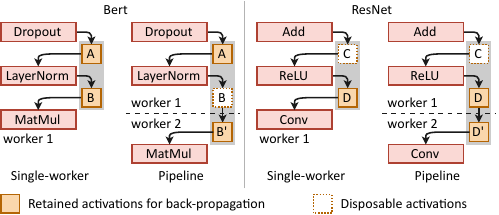}
    \caption{The comparison of activation memory consumptions. The memory consumption in ResNet increases due to the doubled memory footprint of tensor D.}
    \label{fig:pipemem}
\end{figure}

In ResNet, the measured memory usage of pipeline parallelism increases with the number of GPUs. Figure \ref{fig:pipemem} demonstrates the reason for the inconsistent memory usage of pipeline parallelism compared to expectations. In Bert, the back-propagation of the LayerNorm operator depends on the input tensor $A$, and the output tensor $B$ is only used for the back-propagation of the next operator. Therefore, in the pipeline parallelism, $B$ can be destroyed after being passed to the next worker. In contrast, as $C$ is not retained after the in-place ReLU computation, the back-propagation of ReLU depends on the output tensor $D$. In the pipeline parallelism, $D$ must be retained after being transmitted to the next worker, causing an increase in memory footprint.

\textbf{Remark12} Placing in-place operators at the end of a stage in pipeline parallelism will result in an increase in memory consumption for activations.

\begin{figure}[t]
    \centering
    \begin{subfigure}[b]{0.49\linewidth}
        \includegraphics[width=\textwidth]{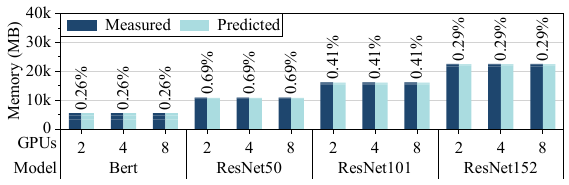}
        \caption{Data parallelism}
        \label{fig:preddatamem}
    \end{subfigure}
    \hfill
    \begin{subfigure}[b]{0.49\linewidth}
        \includegraphics[width=\textwidth]{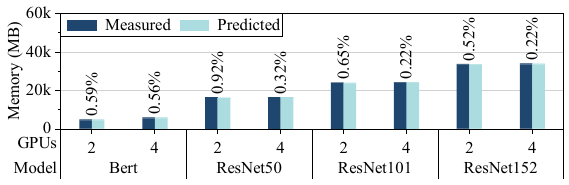}
        \caption{Tensor parallelism}
        \label{fig:predtensormem}
    \end{subfigure}
    \caption{The comparison of measured and predicted memory footprint for data and tensor parallelism, considering the communication buffer and the ``concat'' operator, respectively. The percentage error is shown above each prediction.}
\end{figure}

For data parallelism, the weights consume twice the predicted memory. This is because PyTorch retains the original memory space for weights and creates buffers for weights aggregations when calling DistributedDataParallel. When these buffers are taken into account, our predictions and measurements of the memory overhead for data parallelism align, as shown in Figure \ref{fig:preddatamem}. In terms of tensor parallelism, the implementation of tensor parallelism in ResNet requires the additional ``concat'' operators to concatenate tensors, thereby introducing extra memory overhead. If we take into account the ``concat'' operators, our predictions and measurements of the memory overhead for tensor parallelism align, as illustrated in Figure \ref{fig:predtensormem}.

\textbf{Remark13} The memory overhead during training can be estimated by considering the size of weights and activation tensors.

\section{Discussion}\label{sec:discussion}

\subsection{Key findings}

\begin{table}[t]
\caption{A comparison of distributed deep learning strategies in terms of communication (Comm.) volume, communication pattern, wait in communication, computation (Comp.), overlap, scheduling overhead, memory (Mem.), and suitable optimization.}
\centering
\begin{tabular}{@{}lcccccccc@{}}
\toprule
Strategy & Comm. volume & Comm. pattern & Wait & Comp. & Overlap & Scheduling & Mem. & Optimization \\ \midrule
Data & Weight & Collective & Yes & - & Most & Negligible & ↑ & ZeRO \\
Tensor & Weight/Activation & Collective & Yes & ↑ & Scarce & Negligible & ↑ & / \\
Pipeline & Activation & P2P & No & - & Most & Many & - & Re-computation \\ \bottomrule
\end{tabular}
\label{tab: strategyconclude}
\end{table}

\textbf{Most experimental results agree with our theoretical analysis}, as concluded in Table \ref{tab: strategyconclude}. Data parallelism has typically low data traffic and allows effective overlap of communication and computation. Due to the lack of obvious disadvantages, data parallelism is highly versatile. The scheduling overhead of pipeline parallelism is notably more significant. Even when the number of micro-batches is set to be several times the parallel degree, there is still a substantial amount of device idleness. Tensor parallelism incurs higher communication time, as there is limited overlap between communication and computation, making it challenging to achieve high performance.

We also highlight the below exceptional observations:
\begin{itemize}
    \item The experimental results reveal that the common assumptions about computation time \textbf{significantly differ from reality}. Practically, the total computation time for data and tensor parallelism will increase due to the computational resources occupied by communication, and the computational overhead of the \textit{AR pattern} operators(Remark 6). Recent automatic parallelism methods assume that the total computation time remains unchanged regardless of how the operators are divided, which may introduce deviations.
    \item Placing in-place operators at the end of a pipeline stage \textbf{increases the memory usage of activations}. This is because two operators, which refer to the same tensor for back-propagation, are divided into two different stages. Therefore, designing a pipeline scheme must be concerned about this situation to avoid unnecessary memory overhead (Remark 12).
    \item The main performance bottleneck of pipeline parallelism in PyTorch is \textbf{ineffective communication overlapping}. The pipeline parallelism implementation in PyTorch cannot effectively overlap communication with computation during forward propagation due to unnecessary synchronization (Remark 7), despite the theoretically high communication overlap rate of pipeline parallelism.
\end{itemize}

Additionally, the experimental results indicate that \textbf{the selection of communication block size and reducing waiting time} are further performance optimization directions. Our experiments demonstrated that overlapping cannot completely offset the impact of communication time on scalability (Remark 9). To optimize performance, it is necessary to both reduce communication time and increase the proportion of overlap time during communication. This requires meticulously selecting the size of the data parallel communication blocks (Remark 1), as too small blocks fail to fully utilize the bandwidth, while too large blocks increase the time of the last non-overlapped communication. Besides, high scalability demands design considerations for wait times, as wait times between workers are the main communication performance bottleneck when scaling up within a node (Remark 4).

Finally, our experimental results indicate that \textbf{Transformers are more suitable for pipeline parallelism}. CNN networks, due to their negligible weight sizes, have lower communication and memory overheads for data parallelism. However, due to the different relationships between the sizes of activations and weights, Transformers have smaller data traffic and shorter communication times for pipeline parallelism (Remark 3). On the other hand, in data-parallel training, data transmission overhead will become the main bottleneck for Transformers in bandwidth-limited environments (Remark 4). Moreover, due to the large communication volume, a significant amount of communication overhead cannot be overlapped (Remark 8). Lastly, pipeline parallelism significantly reduces memory overhead in Transformers, due to the non-negligible memory cost of weights, gradients, and optimizer states in Transformers (Remark 11).

\subsection{Study on optimization techniques}

\begin{figure}[t]
    \centering
    \includegraphics[width=\linewidth]{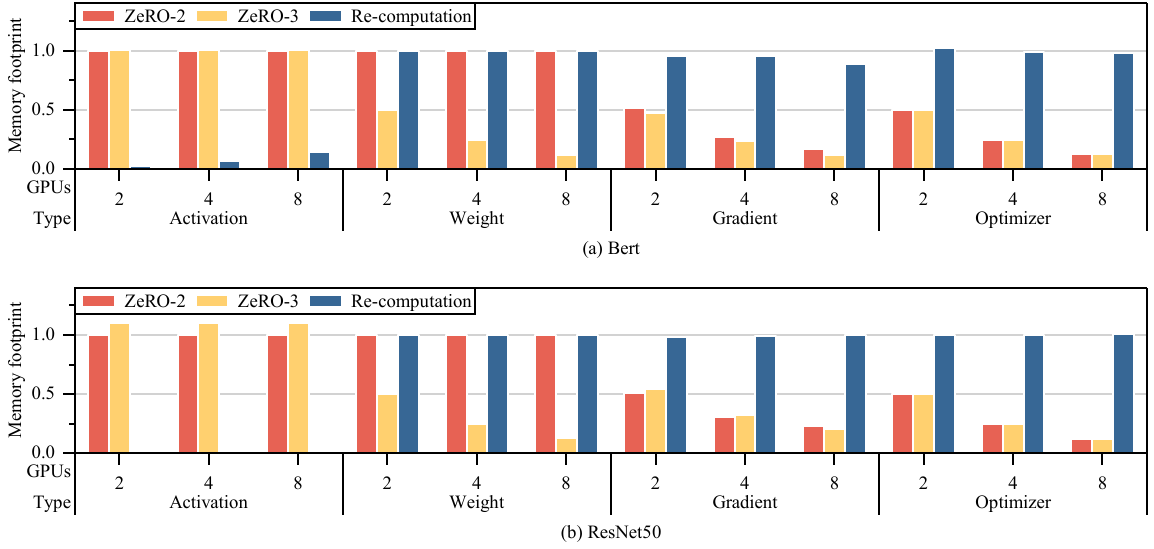}
    \caption{The impact of optimization techniques on different types of memory overhead. All results are divided by the baseline, which is measured without any optimizations.}
    \label{fig:optmem}
\end{figure}

We evaluate the effect of memory optimization techniques through experiments and explain the experimental results based on theoretical analysis.
Figure \ref{fig:optmem} presents the different types of memory overhead when employing DeepSpeed \cite{rasley2020deepspeed} with ZeRO-2 and ZeRO-3 configurations. All results are divided by the baseline, which represents the memory usage without using any optimization techniques. It can be observed that the reduction in the memory usage of gradients and optimizer states for ZeRO-2 is directly proportional to the number of devices. Besides, ZeRO-3 also reduces the memory footprint of weights analogously.

Figure \ref{fig:optmem} also illustrates the memory usage in pipeline parallelism with the adoption of re-computation, which is normalized by the baseline. We can observe that the introduction of re-computation leads to a significant reduction in the memory footprint for activations while keeping other types of memory overhead unchanged.

\textbf{Analysis for ZeRO:} Taking the memory overhead of data parallelism as an example, after using ZeRO-2, the total memory overhead becomes from (\ref{eq:datamemory}) to: 
\begin{equation}
    \sum_{d=1}^D M_{data}^d = (\alpha + D + 1) \sum_{p\in \mathcal{P}} w_p + \sum_{p\in\mathcal{P}} a_p 
    \enskip.
\end{equation}
ZeRO-3 can further partition the weights in \textit{R-layout}, reducing the memory as: 
\begin{equation}
    \sum_{d=1}^D M_{data}^d = (\alpha + 2) \sum_{p\in \mathcal{P}} w_p + \sum_{p\in\mathcal{P}} a_p 
    \enskip.
\end{equation}

\textbf{Analysis for re-computation:} For pipeline parallelism, re-computation only retains the input tensor of the first operator in each stage. For each operator, the sizes of all input tensors become $1/B$ of the original with re-computation, except for the first operator of each stage. Taking the lower bound of pipeline parallelism, GPipe, as an example, the memory overhead with re-computation becomes from (\ref{eq:pipememory}) to
\begin{equation}
    \sum_{d=1}^D M_{GPipe}^{d} = (\alpha+2)\sum_{p\in\mathcal{P}} w_p +  \sum_{p\in\mathcal{P}} \frac{a_p}{B} + (B-1)\sum_{d=1}^D \frac{a_{p(d,1)}}{B} 
    \enskip.
\end{equation}
We further analyze whether recalculation can be used for other distributed strategies. Ideally, if data and tensor parallelism retain $D$ activations consistently with pipeline parallelism, and these $D$ activations evenly divide the memory overhead of the activation values into $D$ parts, then the total memory overhead of the other activations becomes $1/D$ of the original. As our analysis in scheduling overhead, $B$ is set to a multiple of $D$, as in GPipe where $B=4D$. Therefore, re-computation can be used more efficiently with pipeline parallelism. 

\section{Related work}\label{sec:related}

\textbf{Distributed training performance modeling.}
Research on performance modeling can provide support for the scheduling and automatic deployment of training tasks in hardware platforms. These performance modeling methods can be mainly divided into three categories: analytical models, simulator-based models, and prediction models based on machine learning methods. Analytical models, such as PALEO \cite{qi2017paleo} and DistIR \cite{Santhanam2021DistIR}, use the floating-point operation counts and hardware compute capability to evaluate the computation time. Simulator-based models, such as Daydream \cite{Zhu2020Daydream}, dPRO \cite{Hu2022dpro}, and DistSim \cite{Lu2023DistSim}, actually run the model to obtain the execution time of each kernel and simulate the training process. Machine learning prediction models, such as Habitat \cite{Yu2021Habitat} and the method proposed by Wu et al. \cite{Wu2022Machine}, use multi-layer perceptrons to predict the computation time of operators with specified input and output tensor shapes.

\begin{table}[tb]
\caption{Comparison between works for performance modeling.}
\label{tab:related-work-modeling}
\centering
\begin{tabular}{@{}llll@{}}
\toprule
Methodology         & Accuracy & Zero running & Interpretability\\ \midrule
Analytical model \cite{qi2017paleo, Santhanam2021DistIR} 
                    & \ding{55} & \ding{51}     & \ding{51}         \\
Simulator \cite{Zhu2020Daydream, Hu2022dpro, Lu2023DistSim}
                    & \ding{51} & \ding{55}     & \ding{51}         \\
ML model \cite{Yu2021Habitat, Wu2022Machine}
                    & \ding{51} & \ding{51}     & \ding{55}         \\
Ours                & \ding{51} & \ding{51}     & \ding{51}         \\ \bottomrule
\end{tabular}
\end{table}

It is worth noting that our goal is to theoretically analyze the performance of Transformers and traditional networks under different distributed strategies, rather than to predict training performance. Compared to previous work, our analysis simultaneously possesses the characteristics of analytical models and simulator-based models, as in Table \ref{tab:related-work-modeling}. For example, we compute the communication time of different distributed strategies through theoretical bandwidth to avoid the profiling process. Besides, when analyzing the overlap of computation and communication time and scheduling overhead, we treat the execution time and communication time of each operator in the model as known, avoiding strong assumptions. In addition, we also consider the additional communication costs and scheduling overhead introduced by tensor parallelism and pipeline parallelism, which are overlooked in previous simulation-based modeling.

\textbf{Evaluation of deep learning training.}
An increasing number of endeavors are being made to address training performance issues through evaluation, such as Fathom \cite{adolf2016fathom}, TBD \cite{zhu2018benchmarking}, and DawnBench \cite{coleman2019dawnbench}. MLPerf \cite{mattson2020mlperftraining, mattson2020mlperf} further proposes a time-to-train metric to combine accuracy and performance metrics for more practical conclusions. Some works mainly focus on the performance of deep learning frameworks \cite{kim2017performance, Lu2023Quantitative} and computing libraries \cite{shi2016benchmarking}. This article further demonstrates the effect of distributed strategies on training from the perspectives of computation, communication, and memory footprint.

\begin{table}[t]
\centering
\caption{Comparison between recent distributed training performance evaluation studies in terms of distributed strategies, optimization, training time, memory footprint, and used models.}
\label{tab:related-work-evaluation}
\begin{tabular}{@{}lllllll@{}}
\toprule
Work & Strategy & Optimization & Time & Memory & CNN & Transformer \\ \midrule
\cite{shams2017evaluation} & Data parallelism & \ding{55} & \ding{51} & \ding{55} & \ding{51} & \ding{55} \\
\cite{shi2018dag} & Data parallelism & \ding{55} & \ding{51} & \ding{55} & \ding{51} & \ding{55} \\
\cite{mahon2020performance} & Data parallelism & \ding{55} & \ding{51} & \ding{55} & \ding{51} & \ding{55} \\
\cite{ko2021depth} & Data parallelism & \ding{51} & \ding{51} & \ding{55} & \ding{51} & \ding{55} \\
\cite{krisilias2021performance} & Data parallelism & \ding{55} & \ding{51} & \ding{55} & \ding{51} & \ding{55} \\
\cite{ponnuswami2022evaluating} & Data parallelism & \ding{55} & \ding{51} & \ding{51} & \ding{51} & \ding{55} \\
Ours & Data, tensor, and pipeline parallelism & \ding{51} & \ding{51} & \ding{51} & \ding{51} & \ding{51} \\ \bottomrule
\end{tabular}
\end{table}

The impact introduced by distributed training is a focus of research. Sham et al. \cite{shams2017evaluation} evaluated the time and throughput of TensorFlow and Caffe on scale-up and scale-out computing platforms. Shi et al. \cite{shi2018dag} evaluated Caffe, CNTK, MXNet, and TensorFlow using two clusters by breaking down the training time into different stages. Mahon et al. \cite{mahon2020performance} evaluated the scalability of several deep learning frameworks on training using data parallelism. Ko et al. \cite{ko2021depth} analyzed seven centralized and decentralized data-parallel distributed training algorithms from the aspects of model accuracy, hyperparameter sensitivity, scalability, and the effect of optimization techniques. Krisilias et al. \cite{krisilias2021performance} further analyzed the performance of deep learning training on CPU clusters by breaking down time into stages. Ponnuswami et al. \cite{ponnuswami2022evaluating} evaluate the performance of distributed training in synchronous and asynchronous data-parallel settings using TensorFlow.
Compared with these works, as in Table \ref{tab:related-work-evaluation}, this article conducts a holistic evaluation incorporating data, tensor, and pipeline parallelism to analyze the computation, communication, overlapping time, scheduling overhead, and memory footprint.

\textbf{Perspective of distributed strategies.}
To automate the deployment of distributed training tasks, many works abstracted the model partitioning in distributed training strategies. Mesh-TensorFlow \cite{shazeer2018mesh} and GShard \cite{lepikhin2021gshard} proposed a concept to specify the mapping of tensor dimensions to device mesh dimensions, thus treating data parallelism and tensor parallelism as a unified distributed strategy. OneFlow \cite{yuan2021oneflow} proposed the SBP abstraction to describe how input tensors of operators are distributed across different devices, thereby automatically deriving the SBP of output tensors and the required communication for operator computation. Alpa \cite{zheng2022alpa} categorized distributed strategies into intra-operator parallelism and inter-operator parallelism and introduced sharding spec in intra-operator parallelism to define the layout of a tensor, thereby inferring the communication and re-sharding overhead required for each intra-operator parallel algorithm.

However, existing works for tensor parallelism are to sweep all possible parallel algorithms for primitive operators. This approach can be used for performance modeling and the search for distributed training strategies, but it also requires a predefined model structure as input. In comparison to previous work, we further classify the distributed strategies based on operator layout and tensor pattern, thus constructing a more general performance analysis framework.

\section{Conclusions}\label{sec:conclusion}
This paper compares the performance of Transformers and conventional models under different distributed strategies through analytical assessment and experimental comparison. Based on four observations related to model architectures, distributed strategies, and resource consumption, this paper proposes an analytical framework to analyze performance and memory overhead in terms of the tensor sizes and computation times of models, as well as the operator and tensor layouts across devices. Based on the analytical framework, this paper analyzes the training performance and memory overhead for distributed training. The extensive experimental examination corroborates the analytical insights and unveils performance issues left to be addressed. The experimental results also indicate the performance differences between Transformers and conventional models, proving the effectiveness of pipeline parallelism for Transformer model training.

\bibliographystyle{unsrt}  
\bibliography{references} 

\begin{thebibliography}{10}

\bibitem{krizhevsky2017imagenet}
Alex Krizhevsky, Ilya Sutskever, and Geoffrey~E Hinton.
\newblock Imagenet classification with deep convolutional neural networks.
\newblock {\em Communications of the ACM}, 60(6):84--90, 2017.

\bibitem{ren2015faster}
Shaoqing Ren, Kaiming He, Ross Girshick, and Jian Sun.
\newblock Faster r-cnn: Towards real-time object detection with region proposal networks.
\newblock {\em Advances in neural information processing systems}, 28, 2015.

\bibitem{achiam2023gpt}
Josh Achiam, Steven Adler, Sandhini Agarwal, Lama Ahmad, Ilge Akkaya, Florencia~Leoni Aleman, Diogo Almeida, Janko Altenschmidt, Sam Altman, Shyamal Anadkat, et~al.
\newblock Gpt-4 technical report.
\newblock {\em arXiv preprint arXiv:2303.08774}, 2023.

\bibitem{wang2023image}
Wenhui Wang, Hangbo Bao, Li~Dong, Johan Bjorck, Zhiliang Peng, Qiang Liu, Kriti Aggarwal, Owais~Khan Mohammed, Saksham Singhal, Subhojit Som, et~al.
\newblock Image as a foreign language: Beit pretraining for vision and vision-language tasks.
\newblock Proceedings of the IEEE/CVF Conference on Computer Vision and Pattern Recognition, pages 19175--19186. CVF, 2023.

\bibitem{hong2023metagpt}
Sirui Hong, Xiawu Zheng, Jonathan Chen, Yuheng Cheng, Jinlin Wang, Ceyao Zhang, Zili Wang, Steven Ka~Shing Yau, Zijuan Lin, Liyang Zhou, et~al.
\newblock Metagpt: Meta programming for multi-agent collaborative framework.
\newblock {\em arXiv preprint arXiv:2308.00352}, 2023.

\bibitem{osti_1598812}
Ang Li, Shuaiwen Song, Jieyang Chen, Jiajia Li, Xu~Liu, Nathan~R. Tallent, and Kevin~J. Barker.
\newblock Evaluating modern gpu interconnect: Pcie, nvlink, nv-sli, nvswitch and gpudirect.
\newblock {\em IEEE Transactions on Parallel and Distributed Systems}, 31(1), 1 2020.

\bibitem{huang2019gpipe}
Yanping Huang, Youlong Cheng, Ankur Bapna, Orhan Firat, Dehao Chen, Mia Chen, HyoukJoong Lee, Jiquan Ngiam, Quoc~V Le, Yonghui Wu, et~al.
\newblock Gpipe: Efficient training of giant neural networks using pipeline parallelism.
\newblock {\em Advances in neural information processing systems}, 32, 2019.

\bibitem{harlap2018pipedream}
Aaron Harlap, Deepak Narayanan, Amar Phanishayee, Vivek Seshadri, Nikhil Devanur, Greg Ganger, and Phil Gibbons.
\newblock Pipedream: Fast and efficient pipeline parallel dnn training.
\newblock {\em arXiv preprint arXiv:1806.03377}, 2018.

\bibitem{yuan2021oneflow}
Jinhui Yuan, Xinqi Li, Cheng Cheng, Juncheng Liu, Ran Guo, Shenghang Cai, Chi Yao, Fei Yang, Xiaodong Yi, Chuan Wu, et~al.
\newblock Oneflow: Redesign the distributed deep learning framework from scratch.
\newblock {\em arXiv preprint arXiv:2110.15032}, 2021.

\bibitem{mahon2020performance}
Sean Mahon, S{\'e}bastien Varrette, Valentin Plugaru, Fr{\'e}d{\'e}ric Pinel, and Pascal Bouvry.
\newblock Performance analysis of distributed and scalable deep learning.
\newblock 2020 20th IEEE/ACM International Symposium on Cluster, Cloud and Internet Computing (CCGRID), pages 760--766. IEEE, 2020.

\bibitem{shoeybi2019megatron}
Mohammad Shoeybi, Mostofa Patwary, Raul Puri, Patrick LeGresley, Jared Casper, and Bryan Catanzaro.
\newblock Megatron-lm: Training multi-billion parameter language models using model parallelism.
\newblock {\em arXiv preprint arXiv:1909.08053}, 2019.

\bibitem{narayanan2021efficient}
Deepak Narayanan, Mohammad Shoeybi, Jared Casper, Patrick LeGresley, Mostofa Patwary, Vijay Korthikanti, Dmitri Vainbrand, Prethvi Kashinkunti, Julie Bernauer, Bryan Catanzaro, et~al.
\newblock Efficient large-scale language model training on gpu clusters using megatron-lm.
\newblock Proceedings of the International Conference for High Performance Computing, Networking, Storage and Analysis (SC '21), pages 1--15. ACM, 2021.

\bibitem{zheng2022alpa}
Lianmin Zheng, Zhuohan Li, Hao Zhang, Yonghao Zhuang, Zhifeng Chen, Yanping Huang, Yida Wang, Yuanzhong Xu, Danyang Zhuo, Eric~P Xing, et~al.
\newblock Alpa: Automating inter- and {Intra-Operator} parallelism for distributed deep learning.
\newblock 16th USENIX Symposium on Operating Systems Design and Implementation (OSDI 22), pages 559--578. USENIX Association, 2022.

\bibitem{unger2022unity}
Colin Unger, Zhihao Jia, Wei Wu, Sina Lin, Mandeep Baines, Carlos Efrain~Quintero Narvaez, Vinay Ramakrishnaiah, Nirmal Prajapati, Pat McCormick, Jamaludin Mohd-Yusof, et~al.
\newblock Unity: Accelerating $\{$DNN$\}$ training through joint optimization of algebraic transformations and parallelization.
\newblock 16th USENIX Symposium on Operating Systems Design and Implementation (OSDI 22), pages 267--284. USENIX Association, 2022.

\bibitem{paszke2019pytorch}
Adam Paszke, Sam Gross, Francisco Massa, Adam Lerer, James Bradbury, Gregory Chanan, Trevor Killeen, Zeming Lin, Natalia Gimelshein, Luca Antiga, et~al.
\newblock Pytorch: An imperative style, high-performance deep learning library.
\newblock {\em Advances in neural information processing systems}, 32, 2019.

\bibitem{sergeev2018horovod}
Alexander Sergeev and Mike Del~Balso.
\newblock Horovod: fast and easy distributed deep learning in tensorflow.
\newblock {\em arXiv preprint arXiv:1802.05799}, 2018.

\bibitem{rasley2020deepspeed}
Jeff Rasley, Samyam Rajbhandari, Olatunji Ruwase, and Yuxiong He.
\newblock Deepspeed: System optimizations enable training deep learning models with over 100 billion parameters.
\newblock Proceedings of the 26th ACM SIGKDD International Conference on Knowledge Discovery \& Data Mining, pages 3505--3506. ACM, 2020.

\bibitem{chen2016training}
Tianqi Chen, Bing Xu, Chiyuan Zhang, and Carlos Guestrin.
\newblock Training deep nets with sublinear memory cost.
\newblock {\em arXiv preprint arXiv:1604.06174}, 2016.

\bibitem{rajbhandari2020zero}
Samyam Rajbhandari, Jeff Rasley, Olatunji Ruwase, and Yuxiong He.
\newblock Zero: Memory optimizations toward training trillion parameter models.
\newblock SC20: International Conference for High Performance Computing, Networking, Storage and Analysis, pages 1--16. IEEE, 2020.

\bibitem{he2016deep}
Kaiming He, Xiangyu Zhang, Shaoqing Ren, and Jian Sun.
\newblock Deep residual learning for image recognition.
\newblock Proceedings of the IEEE conference on computer vision and pattern recognition, pages 770--778. IEEE, 2016.

\bibitem{deng2009imagenet}
Jia Deng, Wei Dong, Richard Socher, Li-Jia Li, Kai Li, and Li~Fei-Fei.
\newblock Imagenet: A large-scale hierarchical image database.
\newblock 2009 IEEE conference on computer vision and pattern recognition, pages 248--255. IEEE, 2009.

\bibitem{devlin2018bert}
Jacob Devlin, Ming-Wei Chang, Kenton Lee, and Kristina Toutanova.
\newblock Bert: Pre-training of deep bidirectional transformers for language understanding.
\newblock {\em arXiv preprint arXiv:1810.04805}, 2018.

\bibitem{wang2018glue}
Alex Wang, Amanpreet Singh, Julian Michael, Felix Hill, Omer Levy, and Samuel~R Bowman.
\newblock Glue: A multi-task benchmark and analysis platform for natural language understanding.
\newblock {\em EMNLP 2018}, page 353, 2018.

\bibitem{adolf2016fathom}
Robert Adolf, Saketh Rama, Brandon Reagen, Gu-yeon Wei, and David Brooks.
\newblock Fathom: reference workloads for modern deep learning methods.
\newblock 2016 IEEE International Symposium on Workload Characterization (IISWC), pages 1--10. IEEE, 2016.

\bibitem{zhu2018benchmarking}
Hongyu Zhu, Mohamed Akrout, Bojian Zheng, Andrew Pelegris, Anand Jayarajan, Amar Phanishayee, Bianca Schroeder, and Gennady Pekhimenko.
\newblock Benchmarking and analyzing deep neural network training.
\newblock 2018 IEEE International Symposium on Workload Characterization (IISWC), pages 88--100. IEEE, 2018.

\bibitem{coleman2019dawnbench}
Cody Coleman, Daniel Kang, Deepak Narayanan, Luigi Nardi, Tian Zhao, Jian Zhang, Peter Bailis, Kunle Olukotun, Chris R{\'e}, and Matei Zaharia.
\newblock Analysis of dawnbench, a time-to-accuracy machine learning performance benchmark.
\newblock {\em ACM SIGOPS Operating Systems Review}, 53(1):14--25, 2019.

\bibitem{mattson2020mlperftraining}
Peter Mattson, Christine Cheng, Gregory Diamos, Cody Coleman, Paulius Micikevicius, David Patterson, Hanlin Tang, Gu-Yeon Wei, Peter Bailis, Victor Bittorf, et~al.
\newblock Mlperf training benchmark.
\newblock {\em Proceedings of Machine Learning and Systems}, 2:336--349, 2020.

\bibitem{mattson2020mlperf}
Peter Mattson, Vijay~Janapa Reddi, Christine Cheng, Cody Coleman, Greg Diamos, David Kanter, Paulius Micikevicius, David Patterson, Guenther Schmuelling, Hanlin Tang, et~al.
\newblock Mlperf: An industry standard benchmark suite for machine learning performance.
\newblock {\em IEEE Micro}, 40(2):8--16, 2020.

\bibitem{kim2017performance}
Heehoon Kim, Hyoungwook Nam, Wookeun Jung, and Jaejin Lee.
\newblock Performance analysis of cnn frameworks for gpus.
\newblock 2017 IEEE International Symposium on Performance Analysis of Systems and Software (ISPASS), pages 55--64. IEEE, 2017.

\bibitem{tao2018benchip}
Jin-Hua Tao, Zi-Dong Du, Qi~Guo, Hui-Ying Lan, Lei Zhang, Sheng-Yuan Zhou, Ling-Jie Xu, Cong Liu, Hai-Feng Liu, Shan Tang, et~al.
\newblock Benchip: Benchmarking intelligence processors.
\newblock {\em Journal of Computer Science and Technology}, 33(1):1--23, 2018.

\bibitem{blott2021qutibench}
Michaela Blott, Nicholas~J Fraser, Giulio Gambardella, Lisa Halder, Johannes Kath, Zachary Neveu, Yaman Umuroglu, Alina Vasilciuc, Miriam Leeser, and Linda Doyle.
\newblock Evaluation of optimized cnns on heterogeneous accelerators using a novel benchmarking approach.
\newblock {\em IEEE Transactions on Computers}, 70(10):1654--1669, 2020.

\bibitem{ko2021depth}
Yunyong Ko, Kibong Choi, Jiwon Seo, and Sang-Wook Kim.
\newblock An in-depth analysis of distributed training of deep neural networks.
\newblock 2021 IEEE International Parallel and Distributed Processing Symposium (IPDPS), pages 994--1003. IEEE, 2021.

\bibitem{shams2017evaluation}
Shayan Shams, Richard Platania, Kisung Lee, and Seung-Jong Park.
\newblock Evaluation of deep learning frameworks over different hpc architectures.
\newblock 2017 IEEE 37th International Conference on Distributed Computing Systems (ICDCS), pages 1389--1396. IEEE, 2017.

\bibitem{shi2018dag}
Shaohuai Shi, Qiang Wang, Xiaowen Chu, and Bo~Li.
\newblock A dag model of synchronous stochastic gradient descent in distributed deep learning.
\newblock 2018 IEEE 24th International Conference on Parallel and Distributed Systems (ICPADS), pages 425--432. IEEE, 2018.

\bibitem{li2019evaluating}
Ang Li, Shuaiwen~Leon Song, Jieyang Chen, Jiajia Li, Xu~Liu, Nathan~R Tallent, and Kevin~J Barker.
\newblock Evaluating modern gpu interconnect: Pcie, nvlink, nv-sli, nvswitch and gpudirect.
\newblock {\em IEEE Transactions on Parallel and Distributed Systems}, 31(1):94--110, 2019.

\bibitem{krisilias2021performance}
Andreas Krisilias, Nikodimos Provatas, Nectarios Koziris, and Ioannis Konstantinou.
\newblock A performance evaluation of distributed deep learning frameworks on cpu clusters using image classification workloads.
\newblock 2021 IEEE International Conference on Big Data (Big Data), pages 3085--3094. IEEE, 2021.

\bibitem{dean2012large}
Jeffrey Dean, Greg Corrado, Rajat Monga, Kai Chen, Matthieu Devin, Mark Mao, Marc'aurelio Ranzato, Andrew Senior, Paul Tucker, Ke~Yang, et~al.
\newblock Large scale distributed deep networks.
\newblock {\em Advances in neural information processing systems}, 25, 2012.

\bibitem{li2020pytorch}
Shen Li, Yanli Zhao, Rohan Varma, Omkar Salpekar, Pieter Noordhuis, Teng Li, Adam Paszke, Jeff Smith, Brian Vaughan, Pritam Damania, et~al.
\newblock Pytorch distributed: Experiences on accelerating data parallel training.
\newblock {\em arXiv preprint arXiv:2006.15704}, 2020.

\bibitem{xu2020automatic}
Yuanzhong Xu, HyoukJoong Lee, Dehao Chen, Hongjun Choi, Blake Hechtman, and Shibo Wang.
\newblock Automatic cross-replica sharding of weight update in data-parallel training.
\newblock {\em arXiv preprint arXiv:2004.13336}, 2020.

\bibitem{Switch2022Fedus}
William Fedus, Barret Zoph, and Noam Shazeer.
\newblock Switch transformers: Scaling to trillion parameter models with simple and efficient sparsity.
\newblock {\em Journal of Machine Learning Research}, 23(120):1--39, 2022.

\bibitem{liu2022funcpipe}
Yunzhuo Liu, Bo~Jiang, Tian Guo, Zimeng Huang, Wenhao Ma, Xinbing Wang, and Chenghu Zhou.
\newblock Funcpipe: A pipelined serverless framework for fast and cost-efficient training of deep learning models.
\newblock {\em Proceedings of the ACM on Measurement and Analysis of Computing Systems}, 6(3):1--30, 2022.

\bibitem{ma2019paddlepaddle}
Yanjun Ma, Dianhai Yu, Tian Wu, and Haifeng Wang.
\newblock Paddlepaddle: An open-source deep learning platform from industrial practice.
\newblock {\em Frontiers of Data and Domputing}, 1(1):105--115, 2019.

\bibitem{abadi2016tensorflow}
Mart{\'\i}n Abadi, Ashish Agarwal, Paul Barham, Eugene Brevdo, Zhifeng Chen, Craig Citro, Greg~S Corrado, Andy Davis, Jeffrey Dean, Matthieu Devin, et~al.
\newblock Tensorflow: Large-scale machine learning on heterogeneous distributed systems.
\newblock {\em arXiv preprint arXiv:1603.04467}, 2016.

\bibitem{huawei2022huawei}
Ltd. Huawei Technologies~Co.
\newblock Huawei mindspore ai development framework.
\newblock Artificial Intelligence Technology, pages 137--162. Springer, 2022.

\bibitem{Lu2023Quantitative}
Zhengxian Lu, Chengkun Du, Yanfeng Jiang, Xueshuo Xie, Tao Li, and Fei Yang.
\newblock Quantitative evaluation of deep learning frameworks in heterogeneous computing environment.
\newblock {\em CCF Trans. High Perform. Comput.}, Sep 2023.

\bibitem{shi2016benchmarking}
Shaohuai Shi, Qiang Wang, Pengfei Xu, and Xiaowen Chu.
\newblock Benchmarking state-of-the-art deep learning software tools.
\newblock 2016 7th International Conference on Cloud Computing and Big Data (CCBD), pages 99--104. IEEE, 2016.

\bibitem{ponnuswami2022evaluating}
Ganesan Ponnuswami, Sriram Kailasam, and Dileep~Aroor Dinesh.
\newblock Evaluating data-parallel distributed training strategies.
\newblock 2022 14th International Conference on COMmunication Systems \& NETworkS (COMSNETS), pages 759--763. IEEE, 2022.

\bibitem{qi2017paleo}
Hang Qi, Evan~R. Sparks, and Ameet Talwalkar.
\newblock Paleo: A performance model for deep neural networks.
\newblock International Conference on Learning Representations. ICLR, 2017.

\bibitem{Santhanam2021DistIR}
Keshav Santhanam, Siddharth Krishna, Ryota Tomioka, Andrew Fitzgibbon, and Tim Harris.
\newblock Distir: An intermediate representation for optimizing distributed neural networks.
\newblock Proceedings of the 1st Workshop on Machine Learning and Systems, page 15–23, New York, NY, USA, 2021. ACM.

\bibitem{Zhu2020Daydream}
Hongyu Zhu, Amar Phanishayee, and Gennady Pekhimenko.
\newblock Daydream: Accurately estimating the efficacy of optimizations for {DNN} training.
\newblock 2020 USENIX Annual Technical Conference (USENIX ATC 20), pages 337--352. USENIX Association, July 2020.

\bibitem{hu2022dpro}
Hanpeng Hu, Chenyu Jiang, Yuchen Zhong, Yanghua Peng, Chuan Wu, Yibo Zhu, Haibin Lin, and Chuanxiong Guo.
\newblock dpro: A generic performance diagnosis and optimization toolkit for expediting distributed dnn training.
\newblock {\em Proceedings of Machine Learning and Systems}, 4:623--637, 2022.

\bibitem{Lu2023DistSim}
Guandong Lu, Runzhe Chen, Yakai Wang, Yangjie Zhou, Rui Zhang, Zheng Hu, Yanming Miao, Zhifang Cai, Li~Li, Jingwen Leng, and Minyi Guo.
\newblock Distsim: A performance model of large-scale hybrid distributed dnn training.
\newblock Proceedings of the 20th ACM International Conference on Computing Frontiers, page 112–122, New York, NY, USA, 2023. ACM.

\bibitem{Yu2021Habitat}
Geoffrey~X. Yu, Yubo Gao, Pavel Golikov, and Gennady Pekhimenko.
\newblock Habitat: A {Runtime-Based} computational performance predictor for deep neural network training.
\newblock 2021 USENIX Annual Technical Conference (USENIX ATC 21), pages 503--521. USENIX Association, July 2021.

\bibitem{Wu2022Machine}
Ruohan Wu, Mingfan Li, Hanxi Li, Tianxiang Chen, Xinghui Tian, Xiaoxin Xu, Bin Zhou, Junshi Chen, and Hong An.
\newblock Machine learning-enabled performance model for dnn applications and ai accelerator.
\newblock 2022 IEEE 24th Int Conf on High Performance Computing \& Communications; 8th Int Conf on Data Science \& Systems; 20th Int Conf on Smart City; 8th Int Conf on Dependability in Sensor, Cloud \& Big Data Systems \& Application (HPCC/DSS/SmartCity/DependSys), pages 25--34. IEEE, 2022.

\bibitem{lepikhin2021gshard}
Dmitry Lepikhin, HyoukJoong Lee, Yuanzhong Xu, Dehao Chen, Orhan Firat, Yanping Huang, Maxim Krikun, Noam Shazeer, and Zhifeng Chen.
\newblock {GS}hard: Scaling giant models with conditional computation and automatic sharding.
\newblock International Conference on Learning Representations. ICLR, 2021.

\bibitem{shazeer2018mesh}
Noam Shazeer, Youlong Cheng, Niki Parmar, Dustin Tran, Ashish Vaswani, Penporn Koanantakool, Peter Hawkins, HyoukJoong Lee, Mingsheng Hong, Cliff Young, et~al.
\newblock Mesh-tensorflow: Deep learning for supercomputers.
\newblock {\em Advances in neural information processing systems}, 31, 2018.

\bibitem{Li2014scaling}
Mu~Li, David~G. Andersen, Jun~Woo Park, Alexander~J. Smola, Amr Ahmed, Vanja Josifovski, James Long, Eugene~J. Shekita, and Bor-Yiing Su.
\newblock Scaling distributed machine learning with the parameter server.
\newblock 11th USENIX Symposium on Operating Systems Design and Implementation (OSDI 14), pages 583--598, Broomfield, CO, October 2014. USENIX Association.

\bibitem{goodfellow2016deep}
Ian Goodfellow, Yoshua Bengio, and Aaron Courville.
\newblock {\em Deep learning}.
\newblock MIT press, 2016.

\bibitem{hochreiter1997long}
Sepp Hochreiter and J{\"u}rgen Schmidhuber.
\newblock Long short-term memory.
\newblock {\em Neural computation}, 9(8):1735--1780, 1997.

\bibitem{ioffe2015batch}
Sergey Ioffe and Christian Szegedy.
\newblock Batch normalization: Accelerating deep network training by reducing internal covariate shift.
\newblock International conference on machine learning, pages 448--456. PMLR, 2015.

\bibitem{Verbraeken2020survey}
Joost Verbraeken, Matthijs Wolting, Jonathan Katzy, Jeroen Kloppenburg, Tim Verbelen, and Jan~S. Rellermeyer.
\newblock A survey on distributed machine learning.
\newblock {\em ACM Comput. Surv.}, 53(2), mar 2020.

\bibitem{wang2019supporting}
Minjie Wang, Chien-chin Huang, and Jinyang Li.
\newblock Supporting very large models using automatic dataflow graph partitioning.
\newblock Proceedings of the Fourteenth EuroSys Conference 2019. ACM, 2019.

\bibitem{kingma2014adam}
Diederik~P Kingma and Jimmy Ba.
\newblock Adam: A method for stochastic optimization.
\newblock {\em arXiv preprint arXiv:1412.6980}, 2014.

\bibitem{loshchilov2018decoupled}
Ilya Loshchilov and Frank Hutter.
\newblock Decoupled weight decay regularization.
\newblock International Conference on Learning Representations. ICLR, 2019.

\bibitem{dryden2019improving}
Nikoli Dryden, Naoya Maruyama, Tom Benson, Tim Moon, Marc Snir, and Brian Van~Essen.
\newblock Improving strong-scaling of cnn training by exploiting finer-grained parallelism.
\newblock 2019 IEEE International Parallel and Distributed Processing Symposium (IPDPS), pages 210--220. IEEE, 2019.

\bibitem{radford2018gpt}
Alec Radford, Karthik Narasimhan, Tim Salimans, Ilya Sutskever, et~al.
\newblock Improving language understanding by generative pre-training.
\newblock 2018.

\bibitem{radford2019gpt2}
Alec Radford, Jeffrey Wu, Rewon Child, David Luan, Dario Amodei, Ilya Sutskever, et~al.
\newblock Language models are unsupervised multitask learners.
\newblock {\em OpenAI blog}, 1(8):9, 2019.

\bibitem{dehghani2023scalingVIT}
Mostafa Dehghani, Josip Djolonga, Basil Mustafa, Piotr Padlewski, Jonathan Heek, Justin Gilmer, Andreas~Peter Steiner, Mathilde Caron, Robert Geirhos, Ibrahim Alabdulmohsin, et~al.
\newblock Scaling vision transformers to 22 billion parameters.
\newblock International Conference on Machine Learning, pages 7480--7512. PMLR, 2023.

\bibitem{touvron2023llama}
Hugo Touvron, Thibaut Lavril, Gautier Izacard, Xavier Martinet, Marie-Anne Lachaux, Timoth{\'e}e Lacroix, Baptiste Rozi{\`e}re, Naman Goyal, Eric Hambro, Faisal Azhar, et~al.
\newblock Llama: Open and efficient foundation language models.
\newblock {\em arXiv preprint arXiv:2302.13971}, 2023.

\bibitem{yuan2024llm}
Zhihang Yuan, Yuzhang Shang, Yang Zhou, Zhen Dong, Chenhao Xue, Bingzhe Wu, Zhikai Li, Qingyi Gu, Yong~Jae Lee, Yan Yan, et~al.
\newblock Llm inference unveiled: Survey and roofline model insights.
\newblock {\em arXiv preprint arXiv:2402.16363}, 2024.

\bibitem{williams2009roofline}
Samuel Williams, Andrew Waterman, and David Patterson.
\newblock Roofline: an insightful visual performance model for multicore architectures.
\newblock {\em Commun. ACM}, 52(4):65–76, apr 2009.

\end{thebibliography}

\end{document}